\documentclass[%
 reprint,
superscriptaddress,
 amsmath,amssymb,
 aps,
pra,
]{revtex4-2}
\usepackage{subcaption}
\usepackage{graphicx}
\usepackage{dcolumn}
\usepackage{bm}
\usepackage{braket}
\usepackage[dvipsnames]{xcolor}
\usepackage{float}
\usepackage{xfrac}
\usepackage{booktabs}
\usepackage{booktabs}
\usepackage{multirow}
\usepackage[justification=justified, format=plain]{caption}
\usepackage{ragged2e}  
\usepackage{amsfonts}
\usepackage{placeins}  
\usepackage{ragged2e}
\usepackage{adjustbox} 
\usepackage{bbold}
\usepackage{float}

\usepackage{algorithm}
\usepackage{algpseudocode}
\usepackage{amsmath,amssymb}
\usepackage{xcolor}

\begin{document}

\preprint{APS/123-QED}

\title{Quantum Filtering and Stabilization of Dissipative Quantum Systems via Augmented Neural Ordinary Differential Equations}

\author{Shahid Qamar}
\affiliation{School of Science, Harbin Institute of Technology, Shenzhen, 518055, China}

\author{Rana Imran Mushtaq}
\affiliation{School of Science, Harbin Institute of Technology, Shenzhen, 518055, China}

\author{Bo Li}
\email{libo@hit.edu.cn}
\affiliation{School of Science, Harbin Institute of Technology, Shenzhen, 518055, China}

\author{Ho-Kin Tang}
\email{denghaojian@hit.edu.cn}
\affiliation{School of Science, Harbin Institute of Technology, Shenzhen, 518055, China}
\affiliation{Shenzhen Key Laboratory of Advanced Functional Carbon Materials Research and Comprehensive Application, Shenzhen 518055, China.}



\begin{abstract}
Modeling open quantum dynamics without full knowledge of the system Hamiltonian or noise model is a key challenge in quantum control and quantum state estimation. We introduce an Augmented Quantum Neural Ordinary Differential Equation (AQNODE) framework that learns quantum trajectories and dissipation parameters directly from partial continuous measurement data. By embedding the system into a latent space evolved via neural ODEs, AQNODE captures both observable and hidden non-Markovian dynamics with temporal smoothness and physical consistency. Our approach integrates weak measurement data to reconstruct qubit states and time-dependent decoherence rates, enabling accurate state prediction and parameter inference without requiring the explicit physical model during inference. Furthermore, we incorporate AQNODE-based feedback control techniques, including proportional-derivative and time-varying linear-quadratic regulator (LQR) strategies, to steer the quantum system toward target states in real time. Extensive numerical simulations demonstrate AQNODE's ability to generalize across system configurations, achieve low prediction errors, and perform robust quantum filtering and control. These results establish AQNODE as a scalable, differentiable, and potentially extendable to experimentally compatible framework for real-time modeling and control of dissipative quantum systems.

\end{abstract}

\maketitle


\section{Introduction}
Estimation and prediction of the dynamics of quantum systems, which have seen significant advancements during recent decades, are essential to the development of quantum technology~\cite{gebhart2023learning, havlivcek2019supervised, petersen2022special,martinez2017quantum,cortez2017rapid}.  Extracting information from unknown quantum systems typically requires a series of measurements and evaluating the outcomes; however, the complexity and scale of quantum data sometimes make manual interpretation infeasible.  Machine learning (ML) provides an effective solution for extracting patterns and accurately modeling quantum dynamics, even in the presence of external noise and experimental imperfections~\cite{liao2024machine,rigo2020machine,lecun2015deep}.  Integrating ML into quantum measurements improves parameter estimates and system identification~\cite{huang2020predicting,ma2025machine}.  Furthermore, ML-based quantum control allows the formulation of optimal control strategies without comprehensive knowledge of system models, resulting in a data-driven approach for accurate quantum manipulation~\cite{petersen2022special,ma2025machine,niu2019universal}.

Despite significant advances, machine learning methods for quantum systems continue to be challenged by difficulties with scalability, noise robustness, interpretability, and integrating physical constraints~\cite{liao2024machine,rigo2020machine,niu2019universal}.
There are often no theoretical foundations for the practical implementation of quantum machine learning algorithms for noisy intermediate-scale quantum (NISQ) that ensure or at least logically imply that they would perform better than their classical equivalents~\cite{preskill2018quantum,callison2022hybrid, bharti2022noisy}. With the potential to surpass the capabilities of the best classical computers, quantum computing is currently another hot topic of research~\cite{wiseman2009quantum,nielsen2010quantum}. A number of NISQ algorithms have been proposed to advance the NISQ era~\cite{domingo2022optimal}, in which a quantum computer may have hundreds of qubits~\cite{ranvcic2023noisy,wiseman1993quantum}. Understanding the noise present in devices is one of the main obstacles to scaling up NISQ devices. Quantum computer-aided design~\cite{kyaw2021quantum} is one attempt at a solution, but it is still unable to fully account for environmental impacts due to the limited availability of NISQ technology. However, accounting for these undesirable environmental effects would require classical exponential computing resources~\cite{iyer2018small}. Open quantum system research is generally crucial for quantum computing and numerous other branches of physics, including non-equilibrium physics~\cite{bastidas2018floquet}, light-matter interaction~\cite{kyaw2015creation, brown2016co}, and many-body phenomena~\cite{de2017dynamics,flannigan2022many,overbeck2016time}.

Open quantum systems pose a fundamentally challenging estimation problem because their dynamics are continuous in time, are modified by dissipation and decoherence arising from environmental coupling, and are often only partially accessible through weak measurement records~\cite{breuer2002theory,schlosshauer2019quantum}. In realistic settings, one usually does not know the full Hamiltonian or dissipative generator with sufficient accuracy, and the observable trajectory is influenced by hidden, time-dependent, environment-dependent quantities. As a result, the task is not only to predict the visible trajectory, but also to reconstruct the qubit state and infer hidden dissipative rates from incomplete measurement data. A central difficulty in the present setting is that the weak-measurement data does not give the full dynamical state of the open system~\cite{4h4b-3xss, korotkov1999continuous}. The visible qubit trajectory is shaped not only by the Bloch-vector coordinates but also by hidden, time-dependent dissipative processes, which are not directly measured.

Recently, deep learning with neural networks has become a leading paradigm in machine learning, and it is also used to address complex problems in physics~\cite{biamonte2017quantum, iten2020discovering, flam2022learning}. 
Neural ordinary differential equations (ODEs), a class of neural networks defined by continuous-time dynamics, have recently been introduced~\cite {chen2018neural,dupont2019augmented}. Since many physical systems are governed by differential equations, neural ODEs offer a natural framework for learning physical systems~\cite{sholokhov2023physics,zakwan2023physically,van2025neural}. Consequently, a standard Neural ODE trained only on the observable Bloch-vector components could learn an effective trajectory map. However, the environmental influence would remain hidden inside the learned vector field. Such a model would not explicitly return the unknown parameters, which are important for interpreting the open-system dynamics and for designing feedback control. \cite{liang2021modeling,sholokhov2023physics}. This is precisely why augmentation is essential because in an open quantum system, the measured Bloch-vector trajectory is not determined only by the instantaneous state coordinates; it is also shaped by time-dependent dissipative processes generated by the environment. Therefore, it enlarges the learned state representation so that hidden dissipative variables can be inferred jointly with the observable quantum trajectory, thereby turning the problem from pure forecasting into simultaneous filtering and system identification in continuous time. 

Neural ordinary differential equations are well suited for open quantum systems because they learn the vector field governing continuous-time evolution, rather than only a discrete input-output map \cite{choi2022learning,chen2022learning}. This makes them especially appropriate for physical systems whose trajectories are generated by differential equations and for situations in which measurement information arrives sequentially through time. In the present setting, this is particularly valuable because weak-measurement records provide only indirect information about the qubit state, while hidden dissipative processes shape the observed dynamics. A Neural ODE framework therefore provides a natural way to encode measurement history into a latent representation and evolve that representation consistently in continuous time.

Motivated by this perspective, we introduce an Augmented Quantum Neural ODE (AQNODE) framework tailored to non-Markovian open quantum systems. AQNODE learns quantum trajectories and dissipative parameters simultaneously from partial observations. In contrast to Chen et al. \cite{choi2022learning}, whose trajectory fitting lacks physical interpretability, AQNODE enables interpretable parameter inference and feedback control. Neural-controlled differential equations are restricted to closed, unitary quantum dynamics and synthetic tasks. At the same time, AQNODE deals with open system dynamics, explicit dissipation, and robust control based on AQNODE, making it directly relevant for experimental quantum systems ~\cite{yang2024neuralcontrolleddifferentialequations}.

The learned filtering framework also provides a natural basis for feedback control. In a partially observed open quantum system, the controller generally cannot access the full quantum state or hidden dissipative parameters directly~\cite{ZHANG20171,mirrahimi2007stabilizing}. Therefore, accurate state reconstruction is a prerequisite for designing reliable control fields. Designing appropriate control fields for this purpose is one of the main challenges in quantum system control~\cite{ansel2024introduction, d2021introduction}.  However, due to the unavoidable internal fluctuations and external disturbances, the interaction between a quantum system and its environment is usually only partially understood in practical laboratory settings. To mitigate the effects of measurement backaction and real-time disturbances, feedback-based quantum control approaches have been developed, particularly those leveraging continuous weak measurements~\cite{PhysRevX.3.021008, PhysRevA.62.012307, ZHANG20171}. Although many classical control strategies perform well under idealized conditions, their efficacy diminishes significantly when the characteristics of the system vary or are unknown. Traditional open-loop methods (e.g., optimal control~\cite{PhysRevLett.118.150503} and Lyapunov-based quantum control~\cite{PhysRevA.86.022321}) rely heavily on complete knowledge of the system’s Hamiltonian and dissipative dynamics, making them less suitable for scenarios with incomplete information.

In realistic experimental settings, state reconstruction and control design are inseparable. Because the system is only partially observed via continuous weak measurements, feedback cannot be designed directly from the full quantum state. Instead, it must rely on a model that infers the relevant state- and environment-dependent quantities from measurement data. In our framework, AQNODE plays this role: it acts as a learned quantum filter that reconstructs the qubit state and predicts hidden dissipative parameters, and these inferred quantities are then used to design proportional-derivative (PD) and time-varying linear quadratic regulator (LQR) feedback laws for stabilization. The synergy between AQNODE-based modeling and physically grounded control results in a robust system capable of stabilizing quantum states under continuously varying and uncertain conditions.

\begin{figure*}
\includegraphics[width=2\columnwidth]{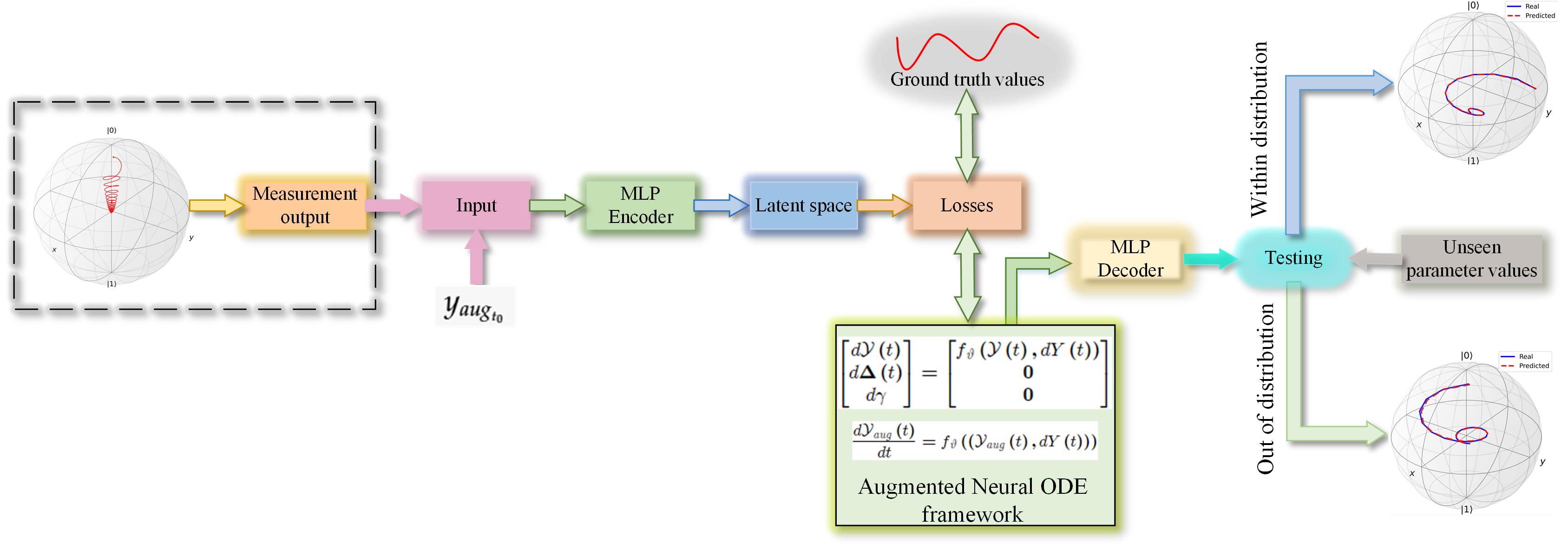}
\caption{\justifying Schematic framework of the AQNODE framework for open quantum state and parameter estimation. The qubit is partially observed through continuous measurement data and is encoded into a latent space together with the initial state, evolved using the Neural ODE solver \( \dot{\mathcal{Y}}_{aug}(t) = f_{\theta}(t, \mathcal{Y}_{aug}(t), dY(t)) \), and a decoder maps the evolved internal state back to physically meaningful quantities. The Adam optimizer is used to minimize the loss \( \mathcal{L} = \sum_i \ell(\mathcal{Y}^i_{aug}(t), \hat{\mathcal{Y}}^i_{aug}(t)) \), which measures the difference between predicted and actual quantum trajectories.}\label{Fig.1}
\end{figure*}

\begin{center}
\textbf{{Summary of results}}
\end{center}
A schematic overview of the framework is shown in Figure~\ref{Fig.1}. The proposed AQNODE framework learns a continuous-time latent representation from weak-measurement data. It decodes it into physically interpretable quantities, including the Bloch vector and the time-dependent dissipative rates. We study a single driven qubit interacting with a dissipative environment. The physical model is a two-level time-convolutionless (TCL) model: the external drive enters through the Hamiltonian, while the non-Markovian environmental effects are represented by structured time-dependent diffusion and dissipation rates obtained from an Ohmic Lorentz-Drude reservoir with finite cutoff \cite{cui2008optimal}.  Instead of assuming the explicit structure of the Hamiltonian or Liouvillian, the model learns an effective neural vector field,
\(\dot{\rho}(t) \approx f_\theta(\rho(t), dY(t), t)\),
that captures both coherent and dissipative contributions to the dynamics.

Our contribution is an end-to-end measurement-driven framework that:
(i) reconstructs the qubit state and hidden dissipative parameters from partial observations;
(ii) turns quantum state estimation into simultaneous system identification in continuous time; and
(iii) uses the AQNODE-inferred state for real-time PD and LQR feedback control to stabilize the system near a desired target state.

We quantify accuracy, physical consistency, and control performance on different performance profiles.  The model is validated in both the within-distribution (WD) and out-of-distribution (OOD) regimes, as well as with initial condition perturbations, displaying significant robustness and generalization. Our framework's scalability and interpretability make it suitable for higher-dimensional and complex quantum systems.  Detailed simulation results (Section IV) demonstrate that AQNODE can reliably reconstruct quantum trajectories and provide feedback in steering the system toward target states under realistic non-Markovian conditions.

\section{System Description}
Open quantum systems are commonly described using quantum master equations, which provide a general framework for modeling non-unitary dynamics induced by coupling to external environments. Unlike isolated quantum systems, whose evolution is governed solely by the Schrödinger equation, open systems experience decoherence and dissipation through interactions with external degrees of freedom. We consider a two-level time-local TCL master equation with a Lindblad-like operator structure in the density-matrix formalism as~\cite{cui2008optimal,qamar2021nonlinear,breuer2016colloquium}
\begin{eqnarray}\label{eqn:1}
\begin{aligned}
\frac{d\rho_t}{dt}&= -\frac{i}{2} \omega_0 [\sigma_z, \rho_t] - \frac{i}{2} u_x(t) [\sigma_x, \rho_t] - \frac{i}{2} u_y(t) [\sigma_y, \rho_t]\\ 
&+ [\Delta(t) + \gamma(t)] \mathcal{D}[\sigma^-] \rho_t +[\Delta(t) - \gamma(t)] \mathcal{D}[\sigma^+] \rho_t\\ &+ M \mathcal{D}[-\frac{\sigma_z}{2}] \rho_t,
\end{aligned}
\end{eqnarray}
where $\omega_0$ is the energy of the system, \(-\frac{i}{2} \omega_0 [\sigma_z, \rho_t]\) describes the unitary evolution due to the qubit's energy difference \(\omega_0\), and \(-\frac{i}{2} u_x(t) [\sigma_x, \rho_t]\) and \(-\frac{i}{2} u_y(t) [\sigma_y, \rho_t]\) represent external controls applied along the \(x\) and \(y\) axes, respectively. The terms $\sigma^-=\frac{1}{2}(\sigma_x-i\sigma_y)$ and $\sigma^+=\frac{1}{2}(\sigma_x+i\sigma_y)$ are the lowering and raising operators, $M\geq0$ is the strength of the system-measurement interaction, and $\mathcal{D}$ denotes the superoperator. \(\mathcal{D}[\sigma^-]\) and \(\mathcal{D}[\sigma^+]\) show dissipative processes such as spontaneous emission and absorption. \(M \mathcal{D}[-\frac{\sigma_z}{2}]\) represents dephasing processes that affect coherence. The superoperator $\mathcal{D}$ is
\begin{eqnarray}\nonumber
\mathcal{D}\left[ L \right]{{\rho }_{t}}=L{{\rho }_{t}}L^{\dagger }-\frac{1}{2}\{L^{\dagger }{L,{\rho }_{t}}\}, \ L \in \{\sigma^-,\sigma^+,\sfrac{\sigma_z}{2}\}
\end{eqnarray}

The parameters $\Delta(t)$ and $\gamma(t)$ are given by~\cite{cui2008optimal, breuer2002theory}.
\begin{align}
\gamma(t)
&=
\frac{\alpha^2\omega_0 r^2}{1+r^2}
\bigg[
1-e^{-r\omega_0 t}
\bigg(
\cos(\omega_0 t)+r\sin(\omega_0 t)
\bigg)
\bigg], \label{eqn:14}
\end{align}
\begin{align}
\Delta(t) & = 2{{\alpha }^{2}}{{k}_{B}}T\frac{{{r}^{2}}}{1+{{r}^{2}}} \nonumber \\ 
& \quad \times \left\{ 1 - {{e}^{-r{{\omega }_{0}}t}} \left[ \cos \left( {{\omega }_{0}}t \right) - \frac{1}{r} \sin \left( {{\omega }_{0}}t \right) \right] \right\}, \label{eqn:16}
\end{align}
where $r$ is the ratio $\omega_c / \omega_0$. The parameters $\Delta(t)$ and $\gamma(t)$ consist of crucial Markovian and non-Markovian characteristics of open quantum systems.  The existence of the environment memory effect is the primary distinction between Markovian and non-Markovian systems.  The environment acts as a sink for information about the quantum system.  Due to the interaction between the system reservoir and the environment, the system loses quantum information.  The non-Markovian dynamics with memory arise from the possibility that the seemingly lost knowledge may return to the system in the future if the environment has a non-trivial structure. The time-dependent rates in Eqns. (2)–(3) follow from the correlation kernels for an Ohmic bath (see Appendix B for more details). Generally, for one qubit, which is a two-dimensional (level) system, we denote $\ket{1}=\begin{pmatrix}
0 \\
1
\end{pmatrix}$ and $\ket{0}=\begin{pmatrix}
1 \\
0
\end{pmatrix}$ as the two eigenvectors of $\sigma_z$. The Pauli matrices $\sigma_z$, $\sigma_y$ and $\sigma_x$ are defined as
\begin{eqnarray}
\sigma_z=\begin{pmatrix}
1     & 0\\
0 & -1
\end{pmatrix},
\ \sigma_y=\begin{pmatrix}
0     & -i\\
i & 0
\end{pmatrix},
\ \sigma_x=\begin{pmatrix}
0     & 1\\
1 & 0
\end{pmatrix}. \nonumber
\end{eqnarray}
The density matrix $\rho_t$ of a two-level quantum system can also be defined by $(x(t),y(t),z(t))$ in the Cartesian coordinate system as
\begin{eqnarray}\label{eqn:4}
{{\rho }{(t)}}=\frac{1}{2}\left( I+{{x}{(t)}}{{\sigma }_{x}}+{{y}{(t)}}{{\sigma }_{y}}+{{z}{(t)}}{{\sigma }_{z}} \right)\nonumber\\ 
=\frac{1}{2}\left( \begin{matrix}
1+{{z}{(t)}} & {{x}{(t)}}-i{{y}{(t)}}  \\
{{x}{(t)}}+i{{y}{(t)}} & 1-{{z}{(t)}}  \\
\end{matrix} \right),
\end{eqnarray}
where $x(t),y(t)$ and $z(t)$ are real numbers, with $tr(\rho_t )=1$, and $tr(\rho_t^2)\leqslant1$.

The ensemble dynamics of the Bloch vector components are given by the following ODEs:
\begin{eqnarray}
\begin{aligned}  
&\frac{dx(t)}{dt} = -\left(\Delta(t) + \frac{M}{2}\right) x(t) - \omega_0 y(t) + u_y(t) z(t)\\
&\frac{dy(t)}{dt} = -\left(\Delta(t) + \frac{M}{2}\right) y(t) + \omega_0 x(t) - u_x(t) z(t)\\
&\frac{dz(t)}{dt} = -2\gamma(t) - 2 \Delta(t) z(t) - u_y(t) x(t) + u_x(t) y(t),
\end{aligned}\label{bloch_eqn}
\end{eqnarray}
where \({\mathcal{Y}(t)}={{\left[ {{x}{(t)}} \ {{y}{(t)}}\ {{z}{(t)}} \right]}^{T}}\in \left\{ {\mathcal{Y}(t) }\in {{\mathbb{R}}^{n}} \mid || \mathcal{Y}(t)  ||^2 \le  1  \right\}\) denotes the Bloch vector representation of the density matrix $\rho_t$. Parameters $\Delta(t)$ and $\gamma(t)$ are time-dependent parameters that depict the diffusion and dissipation coefficients of the quantum system, and \(-\omega_0 y(t)\) and \(\omega_0 x(t)\) describe the precession of the qubit's state vector around the \(z\)-axis. The coefficients \(\Delta(t) + \frac{M}{2}\) and \(2\gamma(t)\) represent the damping effects on the respective components due to environmental interactions.

In matrix notation, the system's dynamics are governed by:
\begin{equation}
\frac{d\mathcal{Y}(t) }{dt} = A(t)\mathcal{Y}(t)  + u_x(t)A_x\mathcal{Y}(t)  + u_y(t)A_y\mathcal{Y}(t) + A_0(t),
\end{equation}
where \(A(t)\) is a time-dependent state matrix describing the system's natural dynamics, \(A_0(t)\) is an inhomogeneous dissipative contribution,  and the matrices \(A_x\) and \(A_y\) define how the control fields \(u_x(t), u_y(t)\) affect the system's evolution; these matrices are defined as
\[
A(t) =
\begin{bmatrix}
-\left(\Delta(t) + \frac{M}{2}\right) & -\omega_0 & 0 \\
\omega_0 & -\left(\Delta(t) + \frac{M}{2}\right) & 0 \\
0 & 0 & -2\Delta(t)
\end{bmatrix},
\]
\[
   A_x =
   \begin{bmatrix}
   0 & 0 & 0 \\
   0 & 0 & -1 \\
   0 & 1 & 0
   \end{bmatrix},
   A_y =
   \begin{bmatrix}
   0 & 0 & 1 \\
   0 & 0 & 0 \\
   -1 & 0 & 0
   \end{bmatrix},
    A_0(t) =
   \begin{bmatrix}
   0 \\
   0 \\
   -2 \gamma(t)
   \end{bmatrix}..
   \]

A photodetector known as a homodyne detector is used to detect the output of the interacting system when a laser probe (along the z-axis) interacts with the atomic ensemble in the weak measurement.  The measured observable is $\sigma_z$ in the z-direction, which gives a continuous measurement of the spin component along the propagation direction.  The quantum system's output is also characterized as
\begin{eqnarray}
\frac{dY(t)}{dt} =\sqrt{M \zeta} tr(-\sigma_z \rho_t),
\end{eqnarray}
where the output of the observation process is $Y(t)$ and $\zeta$ ($0<\zeta<1$) is the detection efficiency. The aforementioned equation demonstrates that the system's output includes some information about its state, and by measuring the system's output, one can gather information about quantum states.  Consequently, adaptive feedback control of the quantum system can be designed by processing the system's output.

In quantum control tasks, control parameters (\(u_x(t), u_y(t)\)) and dissipative terms (\(\gamma(t), \Delta(t)\)) are crucial in determining the behavior of the system and provide points for measurement or manipulation~\cite{fux2021efficient,hwang2012optimal,wu2022trajectory}.  In situations including quantum computing and quantum information processing, an understanding of this system is crucial for improving control and forecasting of qubit behavior.

\section{Quantum Neural ODE}
Formally, the dynamics of the density matrix is governed by an ODE of the form \( \dot{\rho}(t) = f\big(\rho(t), p, t\big) \), where the dynamical rule \( f(\cdot) \), which depends on a set of parameters \( p\), may be unknown.  
The goal is to learn \( f(\cdot) \) from data. 
If the functional form of \( f\big(\rho(t), p, t\big) \) is known but the parameters \( p \) are unknown, the task reduces to estimating these parameters.
In our setup, the functional form of \( f(\cdot) \) is not prescribed a priori but is instead learned through the neural ODE framework.
Using numerical ODE solvers, the neural ODE evolves the system forward in time from \( t_0 \) to \( t_N \) given an initial quantum state \( \rho(t_0) \), producing predictive trajectories \( {\rho}_{\text{pred}}(t_n) \) that approximate the true quantum dynamics. The parameters of the neural ODE are learned by minimizing the prediction error. This work draws inspiration from the latent Neural ODE framework introduced in~\cite{choi2022learning}, where the dynamics are encoded and evolved in a latent space.

In our model, \( f_\theta \) is a trainable neural network to approximate the time derivative of the state, learning directly from data. This makes it a Neural ODE, an ODE where the derivative function is learned, not predefined. This Neural ODE framework enables the model to learn how environmental parameters and control signals influence qubit state evolution. It approximates system trajectories and infers the effects of unobserved dynamics from partial measurements. The learnable parameters \( \theta \), defining the neural dynamics function \( f_\theta \), which include all weights and biases within the neural dynamics function \( f_\theta \), are efficiently optimized using the adjoint sensitivity method, which enables memory-efficient gradient computation through the ODE solver (see Appendix A).

Since the system is continuously monitored through the \(\sigma_z\) measurement channel, the encoder summarizes the measurement record into a latent representation. A latent ODE then advances a compact internal state meant to approximate the Bloch vector while allowing time-varying dissipation rates. A decoder outputs both the qubit state estimate and the instantaneous rate estimates.

\subsection{ Augmented Neural ODE}




The goal of AQNODE is to estimate the qubit state and the time-dependent dissipative coefficients from the available measurement record. The physical state is represented by the Bloch vector \(\mathcal{Y}(t)=[x(t), y(t), z(t)]^T \), while the hidden environmental influence is represented by \( \Delta(t) \) and \( \gamma(t) \). Since these quantities are not all directly observed, AQNODE introduces an internal state estimate that is evolved continuously in time using a Neural ODE.
To capture these variables, we define an augmented state as~\cite{dupont2019augmented}
\begin{equation}
\mathcal{Y}_{\text{aug}}(t) = [x(t), y(t), z(t), \Delta(t), \gamma(t)]^T.
\end{equation}
This augmentation is not merely a change in notation. It is the mechanism that allows the model to convert a partially observed trajectory-learning problem into a joint system-identification problem. If the model were restricted to the observable Bloch coordinates alone, the effect of hidden environment-dependent dissipation would be absorbed implicitly into the learned dynamics, making latent-parameter estimation difficult. By augmenting the state with parameters, AQNODE explicitly represents the hidden physical variables that govern non-Markovian open-system behavior, enabling simultaneous reconstruction of the state and its dissipative drivers from weak measurements.

\subsubsection{ Dynamics of Augmented System}


Since direct access to \( \mathcal{Y}_{\text{aug}}(t) \) is generally unavailable, we initialize a latent trajectory \( h(t) \in \mathbb{R}^d \) from partial data using an encoder network:
\(h_O(t_0) = \text{Encoder}_\psi\left([{\mathcal{Y}_{aug}(t_0)}, dY(t)_{0:t_k}, u_i(t_{0:t_k})]\right)\), where \( \text{Encoder}_\psi \) is a learnable neural network parameterized by \( \psi \) and \(u(t)=[u_x(t),u_y(t)]\).

\textit{Latent Dynamics:} Once initialized, the latent state evolves via a neural ODE. The subsequent latent representation \(h_O(t)\) of the quantum dynamics is produced by an AQNODE layer, which models the continuous dynamics of the system through an ODE parameterized by an MLP with parameters \(\theta\)
\begin{equation}
\frac{d h_O(t)}{dt} = \text{MLP}_\theta([h_O(t), dY(t), u(t), t]).
\end{equation}
The full latent trajectory is computed by solving the ODE over the interval \( t \in [t_0, t_N] \), i.e.,
\begin{equation}
h_O(t_{1:N}) = \text{ODESolve}(h_O(t_0), \text{MLP}_\theta, t_{0:N}),
\end{equation}
where \(t_{0:N}\) are the time points at which the dynamics are evaluated. The function \(\text{ODESolve}\) computes the latent trajectory of AQNODE \(h_O(t)\) and treats it as an initial value problem.

\textit{Decoding the Latent Dynamics:} To extract physically meaningful outputs, \( h_O(t) \) is decoded into the augmented observable space using a decoder network as
\begin{equation}
\hat{\mathcal{Y}}_{\text{aug}}(t) = \text{Decoder}_\phi(h_O(t)) = [\hat{x}(t), \hat{y}(t), \hat{z}(t), \hat{\Delta}(t), \hat{\gamma}(t)]^T,
\end{equation}
where \(\text{Decoder}_\phi\) is an MLP that maps the latent state to the observable and latent variables. This mapping ensures that the entire trajectory of system dynamics, including both observable and latent variables, is captured in \(h_O(t)\). The decoder ensures that the learned latent space is transformed into meaningful physical quantities.

The model is trained by minimizing the state and parameter loss as
\begin{align}
\mathcal{L}_{\text{s}} &= \!\frac{1}{N} \sum_{i,k} \|[\hat{x}_i(t_k), \hat{y}_i(t_k), \hat{z}_i(t_k)] - [x_i(t_k), y_i(t_k), z_i(t_k)]\|^2
\end{align}

\begin{eqnarray}
\mathcal{L}_{\text{p}} = \frac{1}{N} \sum_{i,k} \|[\hat{\Delta}_i(t_k), \hat{\gamma}_i(t_k)] - [\Delta_i(t_k), \gamma_i(t_k)]\|^2  
\end{eqnarray}
Then the total loss becomes
\begin{equation}
\mathcal{L}_{\text{total}} = \kappa \mathcal{L}_{\text{s}} + \beta \mathcal{L}_{\text{p}}
\end{equation}
where \(\kappa\) and \(\beta\) are the hyperparameters.

Although AQNODE is learned through a supervised learning framework, it utilizes the entire system and parameter trajectories as targets during training. Specifically, its architecture is designed to jointly predict unknown parameters and quantum states from partial measurement records at test time, i.e., data that have not been seen before. In this way, AQNODE predicts $x(t), y(t), z(t), \Delta(t)$ and $\gamma(t)$ based on partial system information, and effectively acts as a learned quantum filter.

\subsubsection{Data Generation}
The training data set is generated by simulating the evolution of a single-qubit system under a non-Markovian open quantum dynamics governed by the dissipative TCL dynamics in the simulated parameter regime~(Eq.~\ref{eqn:1})  over a time interval \( t \in [0, T) \) and with continuous weak measurement \(Y(t)\). For each trajectory, the \( \alpha \), \( r \), \( M \), and \( \omega_0 \) are randomly sampled along with the given arbitrary or randomized initial qubit state \( [x_0, y_0, z_0] \). The system is evolved under time-varying parameters \( \Delta(t) \) and \( \gamma(t) \), computed from analytical expressions based on \( \alpha, r, \omega_0 \), and simultaneously driven by external control fields \( u_x(t) \) and \( u_y(t) \). Each training sample consists of the initial augmented state \( [x_0, y_0, z_0, \Delta(0), \gamma(0)] \), the control inputs \( u_x(t), u_y(t) \) (for the state transfer case), and the complete measurement trace \(Y(t)\), which is used to drive the AQNODE. The corresponding ground truth trajectories for the Bloch state and system parameters \( [x(t), y(t), z(t), \Delta(t), \gamma(t)]^T \) are used as supervised targets during training.








\subsubsection{Training}

The training process ought to be adaptable across a wide range of parameter values, as seen by the random sampling of system parameters during simulation.  While our current implementation utilizes simulated data, the system can be adapted to work with measurement data received directly from experiments, enabling future real-time training or inference as compatible data becomes available.

Our model consists of an encoder, a neural ODE function, and a decoder. The network is trained with approximately 130,000 parameters, consisting of 6 input neurons, 128 hidden neurons, and 5 output neurons. The training process is optimized for efficient computation, utilizing parallel processing to accelerate operations. The dataset consists of the number of training trajectories, with each trajectory generated using randomly sampled system parameters \(\alpha, r, M, \omega_0\) with predefined ranges, and together with randomly initialized initial conditions of state. We used the Runge–Kutta technique to get precise numerical solutions of the governing differential equations.  This approach ensures reliable ground-truth trajectories for training, offering high accuracy and stability.

During training, the model is given simulated quantum trajectories that have been generated using sampled system parameters and initial states randomly. The model is trained in a supervised fashion to predict the quantum state and dissipation parameters at each measurement time step, with the simulation output acting as the ground-truth.  This training setting allows the network to learn a mapping from measurement data and initial conditions to the full underlying system evolution.  In this way, one can take the time-series data as a series of continuous weak measurements made across various times $[0, \ t_N ]$ on the statistical ensemble, which is prepared in an initial quantum state.


\subsubsection{Validation}

After training, the model is evaluated on two different test datasets: within-distribution (WD) testing and out-of-distribution (OOD) testing, with different perturbations in the initial qubit state. The WD test set consists of qubit trajectories generated using parameter ranges similar to the training set, while the OOD test set includes trajectories with parameter values extended beyond the training distribution, testing the model's ability to generalize to new conditions.

For each test trajectory, we visualize the predicted and true quantum state evolution to analyze the model's performance further. Our results indicate that the model achieves an MSE of less than $1 \times 10^{-3}$ on WD test cases, and $1 \times 10^{-2}$ on OOD tests, demonstrating precise estimation within the training distribution. The complete process for the data generation, training, and evaluation workflow of AQNODE is summarized in Algorithm 1.

\begin{algorithm}[H]
\caption{Training and evaluation of AQNODE}
\label{alg:aqnode}
\begin{algorithmic}[1]
\Require Time grid $t_{0:N}=\{t_0,\dots,t_N\}$; simulator for Eqns.~(2)--(3), (5) and Eqn. (7);
sampling distributions for $(\rho_0,\alpha,r,\omega_0,M,\zeta)$; AQNODE modules $\mathrm{Encoder}_{\psi}$, $\mathrm{MLP}_{\theta}$, $\mathrm{Decoder}_{\phi}$;
loss weights $(\kappa,\beta)$; optimizer.
\Ensure Trained parameters $(\psi^\star,\theta^\star,\phi^\star)$; estimates $\hat{\mathcal{Y}}(t_k)$, $\hat{Y}_{\mathrm{aug}}(t_k)$.
\vspace{0.25em}
\Statex \underline{\textbf{Generate datasets}}
\For{$s\in\{\mathrm{train},\mathrm{WD},\mathrm{OOD}\}$}
  \For{$i=1,\dots,N_{\mathrm{traj}}^{(s)}$}
    \State Sample $(\alpha,r,\omega_0,M)$ and $\mathcal{Y}_i(t_0)$ from the $s$-range.
    \State Compute $\Delta_i(t_k),\gamma_i(t_k)$ (Eqns.~(2)--(3)).
    \State  Simulate Eqn. (5); set $\mathcal{Y}_{\mathrm{aug},i}(t_k)=[\mathcal{Y}_i(t_k),\Delta_i(t_k),\gamma_i(t_k)]^T$.
    \State Compute $dY_{i}(t_k)=\sqrt{M\zeta}\,\mathrm{tr}(-\sigma_z\rho_i(t_k))$.
    \State Store $\mathcal{D}_i^{(s)}=\{(dY_{i}(t_k),\mathcal{Y}_{\mathrm{aug},i}(t_k),u_i(t_k))\}_{k=0}^N$.
  \EndFor
\EndFor

\vspace{0.25em}
\Statex \underline{\textbf{Train AQNODE on $\mathcal{D}^{(\mathrm{train})}$}}
\State Initialize $(\psi,\theta,\phi)$.
\For{epoch $=1,\dots,E$}
  \State Sample minibatch $\{\mathcal{D}_i^{(\mathrm{train})}\}_{i=1}^{B}$.
  \For{each trajectory $i$ in the minibatch}
    \State Initialize latent state:
    \[
      h_O(t_0)\leftarrow \mathrm{Encoder}_{\psi}\!\big([\mathcal{Y}_{aug,i}(t_0),\,dY_{i}(t_{0:k}), u_i(t_{0:k})]\big).
    \]
    \For{$k=0,\dots,N-1$}
      \State Propagate latent dynamics:
      
      \(   \dot{h}_O(t)=\mathrm{MLP}_{\theta}\!\big([h_O(t),dY_{i}(t), u_i(t),t]\big)
      \)
      
      and set
      
      \(
      h_O(t_{k+1})\leftarrow \mathrm{ODESolve}\big(h_O(t_k),\mathrm{MLP}_{\theta},t_k{:}t_{k+1}\big).
      \)
      \State Decode:
      
      \(
      \hat{\mathcal{Y}}_{\mathrm{aug},i}(t_{k+1})\leftarrow \mathrm{Decoder}_{\phi}(h_O(t_{k+1})).
      \)
    \EndFor
  \EndFor
  \State Compute losses Eqn (14):
  
  \(
  L_{\mathrm{total}}=\kappa L_s+\beta L_p.
  \)
  \State Update $(\psi,\theta,\phi)$ using Adam through \textsc{ODESolve}.
\EndFor
\State Set $(\psi^\star,\theta^\star,\phi^\star)\leftarrow(\psi,\theta,\phi)$.
\vspace{0.25em}
\Statex \underline{\textbf{Evaluate on WD and OOD}}
\For{$s\in\{\mathrm{WD},\mathrm{OOD}\}$}
  \State Run inference on $\mathcal{D}^{(s)}$ and report MSE for $x,y,z,\Delta,\gamma$.
\EndFor
\end{algorithmic}
\end{algorithm}

\section{Results and Discussions}
The simulation results represent the evolution of the qubit's state over time, highlighting the effects of interaction with the environment and the control fields applied to manipulate the qubit. The numerical simulations are conducted in different distinct phases of quantum regimes to evaluate the performance, generalization, and robustness of the AQNODE. These phases involve different configurations of parameter randomization to systematically assess how well the model adapts to WD and OOD scenarios. Phase 1 and Phase 2 operate without the involvement of control inputs, focusing on preliminary AQNODE. Phase 1 is mainly for evaluating generalization to environmental variability, while Phase 2 is mainly for evaluating robustness to system variability. Phase 3 integrates control signals as inputs, enabling the dynamic system to evolve using AQNODE-based control.




\subsection{Phase 1: Environmental noise}
In the first phase, we investigate the impact of environmental parameter variations while keeping system parameters fixed. Specifically, the system is exposed to different levels of dissipation and coupling strength, but \( M \), \( \omega_0 \) remain constant at \( 0.4 \) and \( 1.0 \), respectively.  

\textit{Training Phase}: The model is trained using randomized environmental parameters $\alpha \in [0.4, 0.7] \ \text{and} \ r \in [0.2, 0.5]$.
  
\textit{Testing in WD Setting}:  
To evaluate how well the trained model generalizes to similar but slightly varied conditions, the environmental parameters are sampled from a slightly narrower range, i.e., $\alpha \in [0.45, 0.65] \ \text{and} \ r \in [0.25, 0.45]$
  
\textit{Testing in OOD Setting}:   
To assess the model's ability to extrapolate beyond its training distribution, we test it under wider environmental variations as $\alpha \in [0.2, 0.8] \ \text{and } r \in [0.1, 0.6]$.

\begin{figure*}[t]
    \centering
    \includegraphics[width=0.75\textwidth]{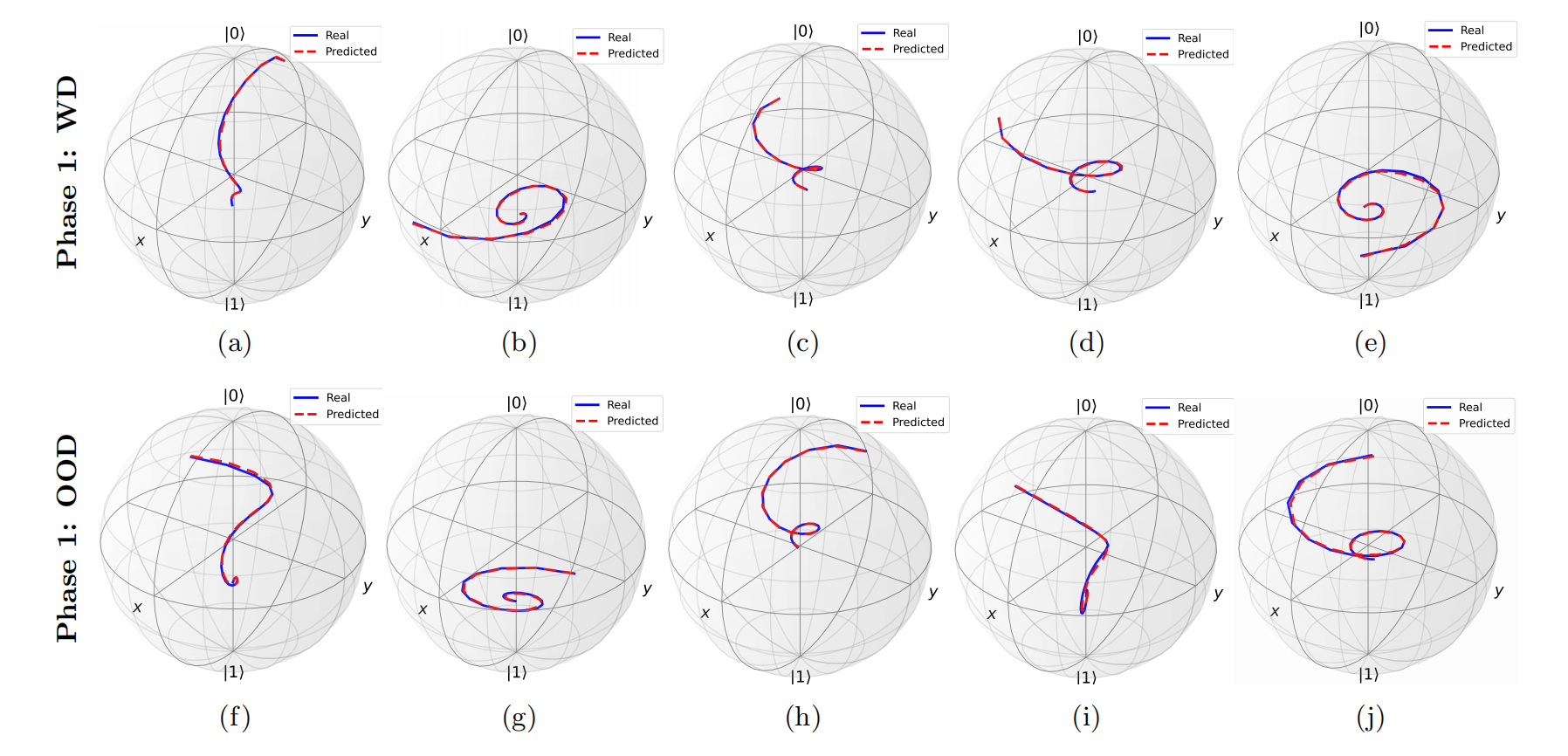}
        \caption{\justifying Augmented AQNODE prediction and evolution of the qubit state in Phase 1 under WD (a–e) and OOD (f–j). The qubit state components \( x(t) \), \( y(t) \), and \( z(t) \) are influenced by the environmental parameters \( r \) and \( \alpha \), which influence the diffusion \( \Delta(t) \) and the dissipation term \( \gamma(t) \). In the WD regime, AQNODE accurately models the quantum state evolution within the training parameter range. In the OOD regime, the model generalizes well, predicting quantum state evolution despite environmental parameters deviate from the training range. Also larger error indicates that extrapolation becomes more difficult when environmental parameters are shifted beyond the training range.}\label{Fig.2}
\end{figure*}

\begin{figure*}[t]
    \centering
    \includegraphics[width=0.75\textwidth]{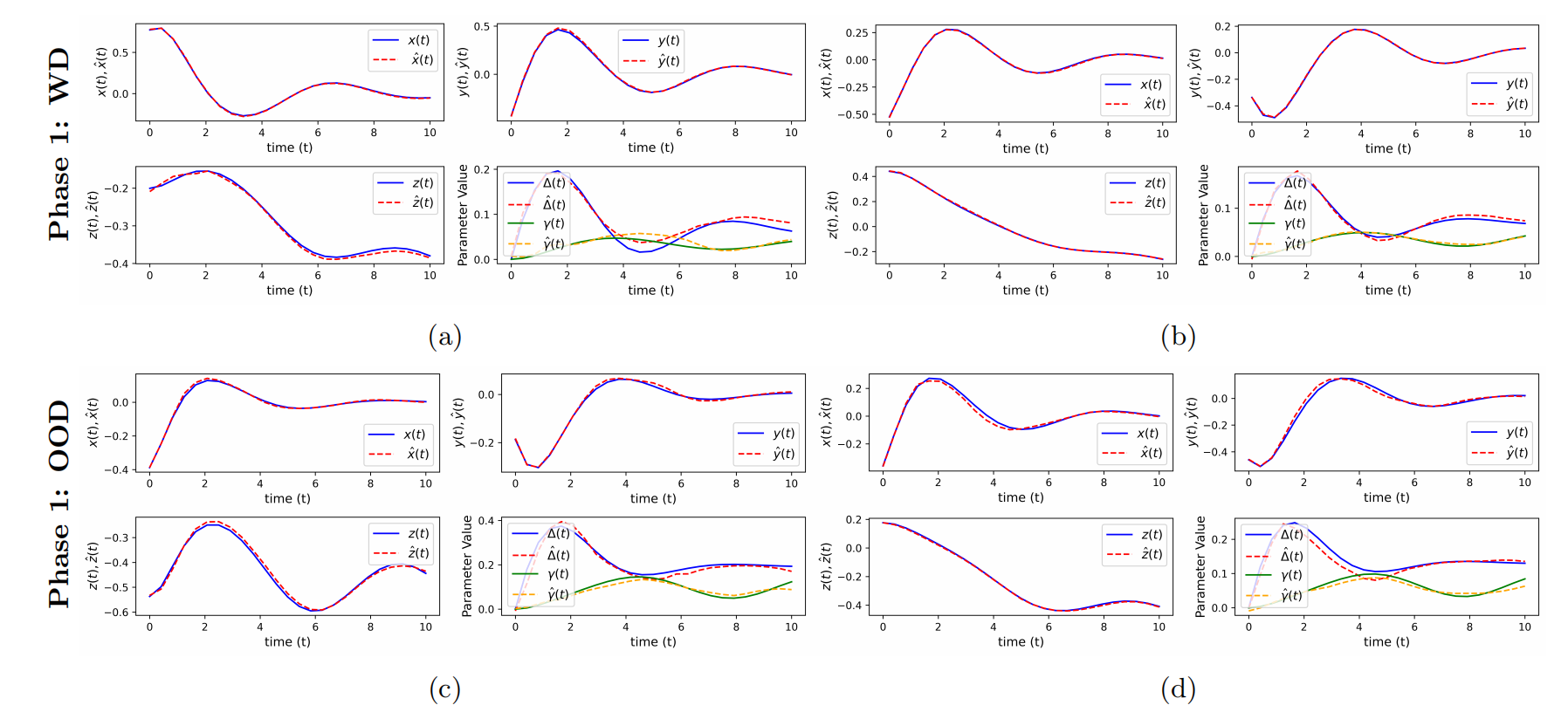}
    \caption{\justifying AQNODE prediction of qubit states and system parameters in Phase 1 under WD (a–b) and OOD (c–d) conditions. The qubit state components \( x(t), y(t), z(t) \) and parameters \( \Delta(t) \) (diffusion) and \( \gamma(t) \) (dissipation) are modeled in the WD scenario, where the environmental parameters \( r \) and \( \alpha \) are within predefined ranges. In the OOD scenario, AQNODE generalizes beyond the training distribution, accurately reconstructing both state dynamics and hidden parameters from measurement outputs, demonstrating the model’s effectiveness in reconstructing both system states and hidden dynamics from limited measurement data.}\label{Fig.3}
\end{figure*}

This phase examines the generalization capacity of AQNODE, ensuring that it can reliably estimate quantum states under different environmental decoherence effects. The evolution of the qubit state is governed by Hamiltonian dynamics, which is directly influenced by the environmental parameters \( r \) and \( \alpha \). These parameters shape the diffusion term \( \Delta(t) \) and the dissipation term \( \gamma(t) \), both of which play a crucial role in the system’s non-Markovian behavior. Visualization is employed to assess model performance, where each trajectory consists of a real quantum trajectory (solid blue curve), representing the ground truth evolution, and the predicted trajectory (dashed red curve), which is generated by integrating the learned Neural ODE dynamics. The performance of AQNODE is measured by how well the expected trajectory aligns with the actual quantum state evolution.

Figures \ref{Fig.2}(a-e) illustrate the phase 1 qubit trajectories under parameter values previously unseen, where variations in \( \alpha \), \( r \), and initial conditions influence the evolution of the system. The AQNODE model generalizes across these conditions and successfully predicts the system dynamics for the WD. The observed trajectories show that the predictions of AQNODE remain closely aligned with trajectories of real qubits, demonstrating its effectiveness in capturing complex quantum evolution governed by dissipation and diffusion effects. Figure~\ref{Fig.2}(f-j) shows the OOD test set of Phase 1, where system parameters deviate from the training range. Although some deviation is visible, the model still captures the qualitative features of quantum evolution, demonstrating partial generalization beyond training conditions. This illustrates both the model's predictive ability in interpolative settings (WD) and its robustness in extrapolative ones (OOD), providing qualitative insight into the learning and generalization behavior of the AQNODE framework in quantum dynamics modeling. The estimation error is very low in both WD and OOD, i.e., less the $1\%$ and $2\%$. To further validate AQNODE's prediction ability over a wider range of conditions, further examples of Bloch sphere trajectories for various test cases are provided.

\begin{figure}[t]
    \centering
    \includegraphics[width=0.4\textwidth]{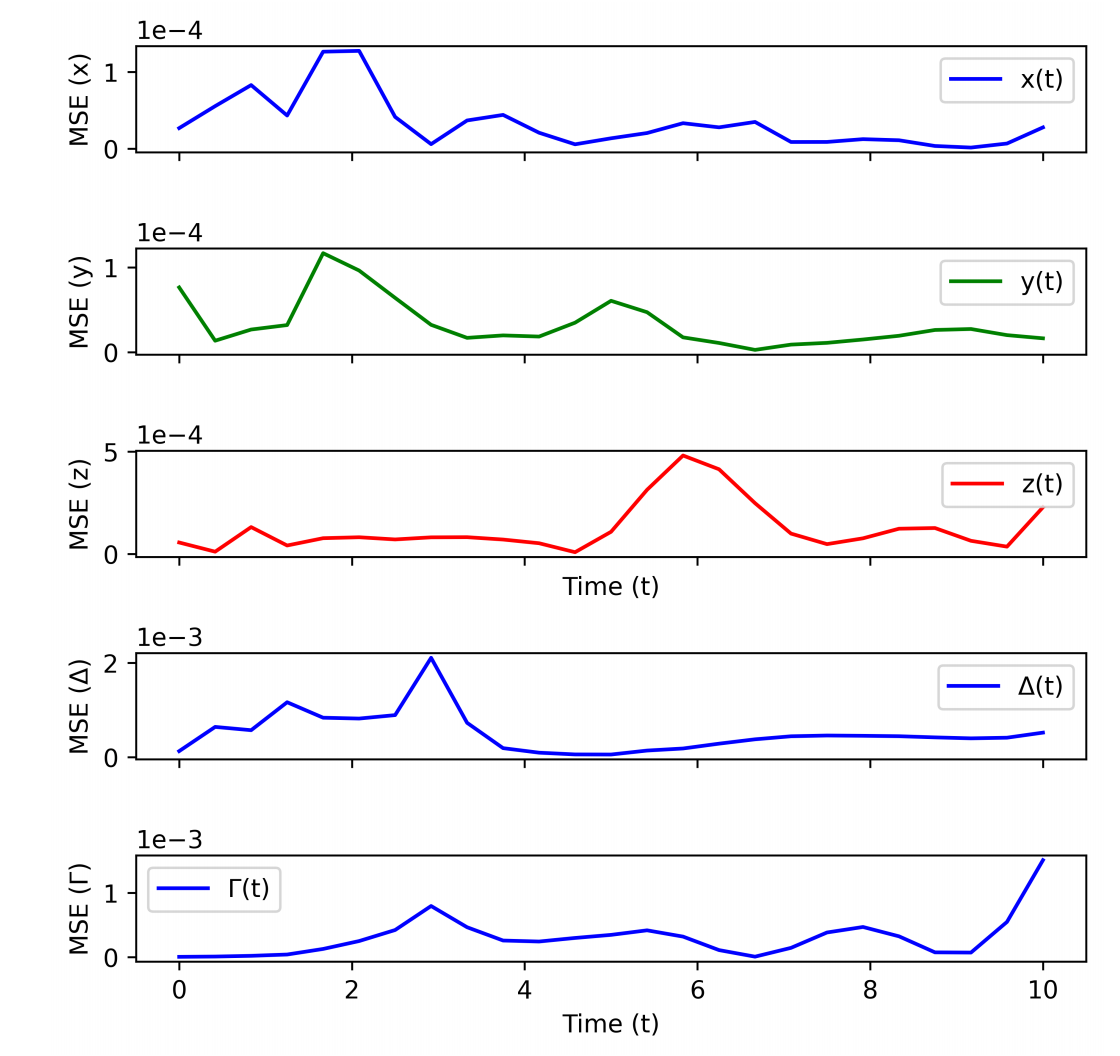}
    \caption{\justifying Time evolution of average MSE during Phase 1 training, showing effective convergence and stable learning of the qubit dynamics from partial measurements.}\label{Fig.4}
\end{figure}

\begin{figure}[t]
    \centering
    \includegraphics[width=0.4\textwidth]{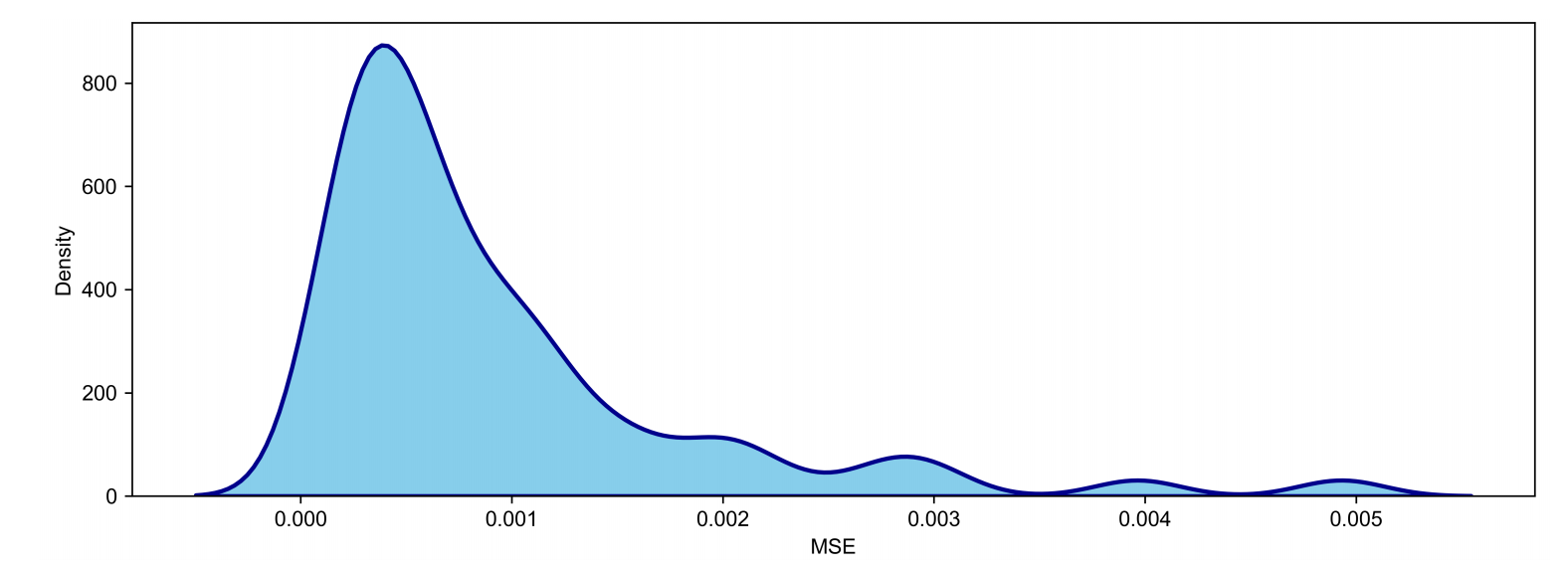}
    \caption{\justifying Phase 1: The distribution exhibits a pronounced peak around MSE \(6\times10^{-4}\), indicating high precision on the majority of training samples. A right-skewed tail suggests challenging trajectories due to complex dynamics or rare configurations, with few high-error outliers.}\label{Fig.5}
\end{figure}

Individual trajectories for each state variable and system parameter predicted during Phase 1 are shown in Figure~\ref{Fig.3}. This figure demonstrates the capability of the AQNODE model to infer both the state evolution and hidden environmental parameters of a quantum system.  Figure~\ref{Fig.3}(a-b) illustrates results from the WD setting, AQNODE shows good predictive estimation, closely matching the ground truth trajectories of the qubit state and the associated environmental dynamics. Furthermore, Figure~\ref{Fig.3}(c-d) shows the performance in the OOD setting, where the test conditions involve parameter values outside the training distribution. Despite this shift, the AQNODE model is effectively generalized, reconstructing the states and latent parameters of the system with reasonable precision. However, WD, the model exhibits a slightly increased prediction error in the OOD regime. This behavior is expected given the inherent distributional mismatch between the training and testing conditions in the OOD regime. The shift in underlying system parameters introduces dynamics that the model has not encountered during training, thereby pushing the limits of its extrapolation capability. Despite this challenge, AQNODE demonstrates robust generalization by accurately predicting both the observable qubit states and the hidden time-dependent parameters. Its augmented state reconstruction mechanism, enables the model to extract meaningful hidden information, even in previously unseen scenarios. This highlights AQNODE’s capacity not only for interpolation within known domains but also for reliable extrapolation beyond the distribution it was trained on. These plots confirm AQNODE’s robustness and its potential to reconstruct quantum states and infer latent system behavior in the dynamical settings where exact parameter distributions may not be known in advance.

\begin{table}[h]
\centering
\caption{\textbf{Phase 1:} Quantitative comparison of MSE for qubit state components and system parameters under WD and OOD.}
\label{tab:mse_summary}
\begin{tabular}{lcc}
\textbf{Metric} & \quad \quad \textbf{WD} & \quad \quad \textbf{OOD} \\
\midrule
MSE~$x(t)$        & \quad \quad \( 2.70  \times 10^{-5} \) & \quad \quad \( 9.80  \times 10^{-5} \) \\
MSE~$y(t)$        & \quad \quad \( 5.35 \times 10^{-5} \) & \quad \quad \( 1.34  \times 10^{-4} \) \\
MSE~$z(t)$        & \quad \quad \( 1.31 \times 10^{-4} \) & \quad \quad \( 6.12 \times 10^{-4} \) \\
MSE~$\Delta(t)$   & \quad \quad \( 2.28 \times 10^{-4} \) & \quad \quad \( 5.83 \times 10^{-4} \) \\
MSE~$\gamma(t)$   & \quad \quad \( 3.17 \times 10^{-4} \) & \quad \quad \( 7.26 \times 10^{-4} \) \\
\bottomrule
\end{tabular}
\label{tab:Phase-1}
\end{table}

Table~\ref {tab:Phase-1} summarizes the MSE performance of the model in predicting the qubit Bloch state components ($x(t)$, $y(t)$, $z(t)$) and system parameters ($\Delta(t)$, $\gamma(t)$) under WD and OOD test settings. The results indicate that the model achieves better performance with minimal error under WD, with low MSE values across all states and parameter estimates. Under OOD settings, performance degrades as expected—particularly for $y(t)$ and $\Delta(t)$—but remains within acceptable error margins, highlighting the model’s reasonable generalization capability even when tested beyond the training distribution.

Figure~\ref{Fig.4} shows an efficient decrease in MSE loss during phase 1. The use of a physics-informed loss function ensured that the model respected the underlying physical laws of the system. Notably, the residuals of the Bloch equations decreased significantly, indicating that the predictions were accurate and physically consistent. Figure~\ref{Fig.5} presents the distribution of MSE values per trajectory on the training dataset, reflecting the model’s prediction precision performance. The sharp peak around MSE $\approx 0.0006-0.0008$ indicates that most trajectories were predicted with high precision, suggesting strong model performance on typical training data. The distribution exhibits a tail on the right that extends toward the MSE $\approx 0.005$, which implies that a smaller subset of trajectories experienced higher prediction errors, likely due to complex dynamics or rare parameter configurations. Minor ripples in the tail may indicate the presence of structurally distinct trajectory classes or slight under-smoothing in the density estimate. This shows that the model generalizes well across the training distribution. 


\subsection{Phase 2: System variation}
In the second phase, we increase the complexity by randomizing both the environmental and system parameters, making the learning task significantly more challenging. This phase tests whether AQNODE can still infer accurate quantum dynamics when the entire system setup varies. 

\begin{figure*}[t]
    \centering
    \includegraphics[width=0.75\textwidth]{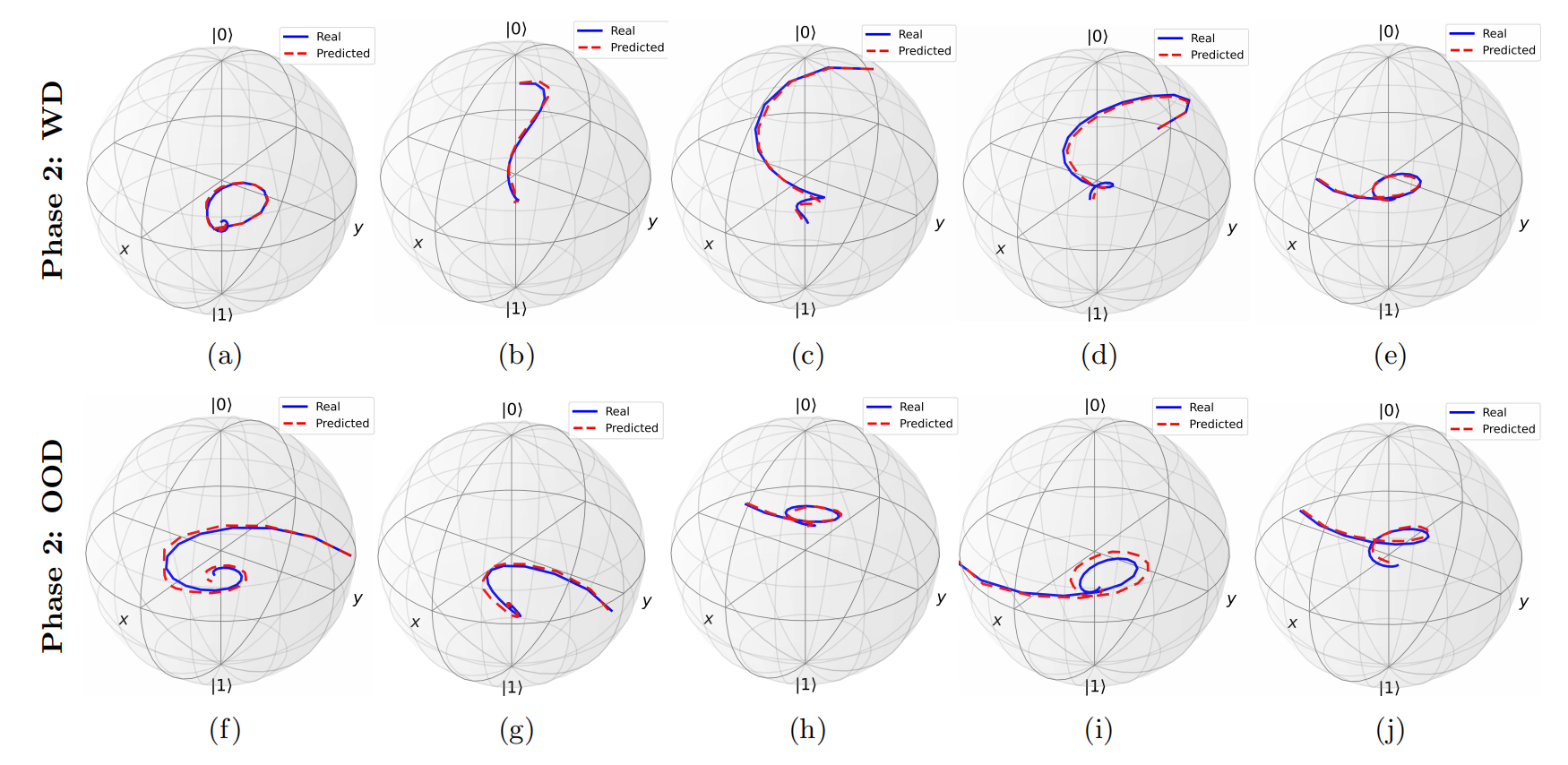}
   \caption{\justifying AQNODE prediction and evolution of the qubit state in Phase 2 under varying non-Markovian and system parameters in WD (a–e) and OOD (f–j). The qubit state variables \( x(t) \), \( y(t) \), and \( z(t) \) are influenced by environmental parameters \( r \) and \( \alpha \), which govern the diffusion \( \Delta(t) \) and \( \gamma(t) \). The qubit's frequency \( \omega_0 \) and measurement strength \( M \) are also varied within the training distribution. Close alignment between true and predicted trajectories shows the model's ability to learn and generalize quantum dynamics. The AQNODE model, trained with limited data, generalizes well to OOD conditions, predicting quantum state evolution despite deviations in environmental parameters.}\label{Fig.6}
\end{figure*}

\begin{figure*}[t]
    \centering
    \includegraphics[width=0.75\textwidth]{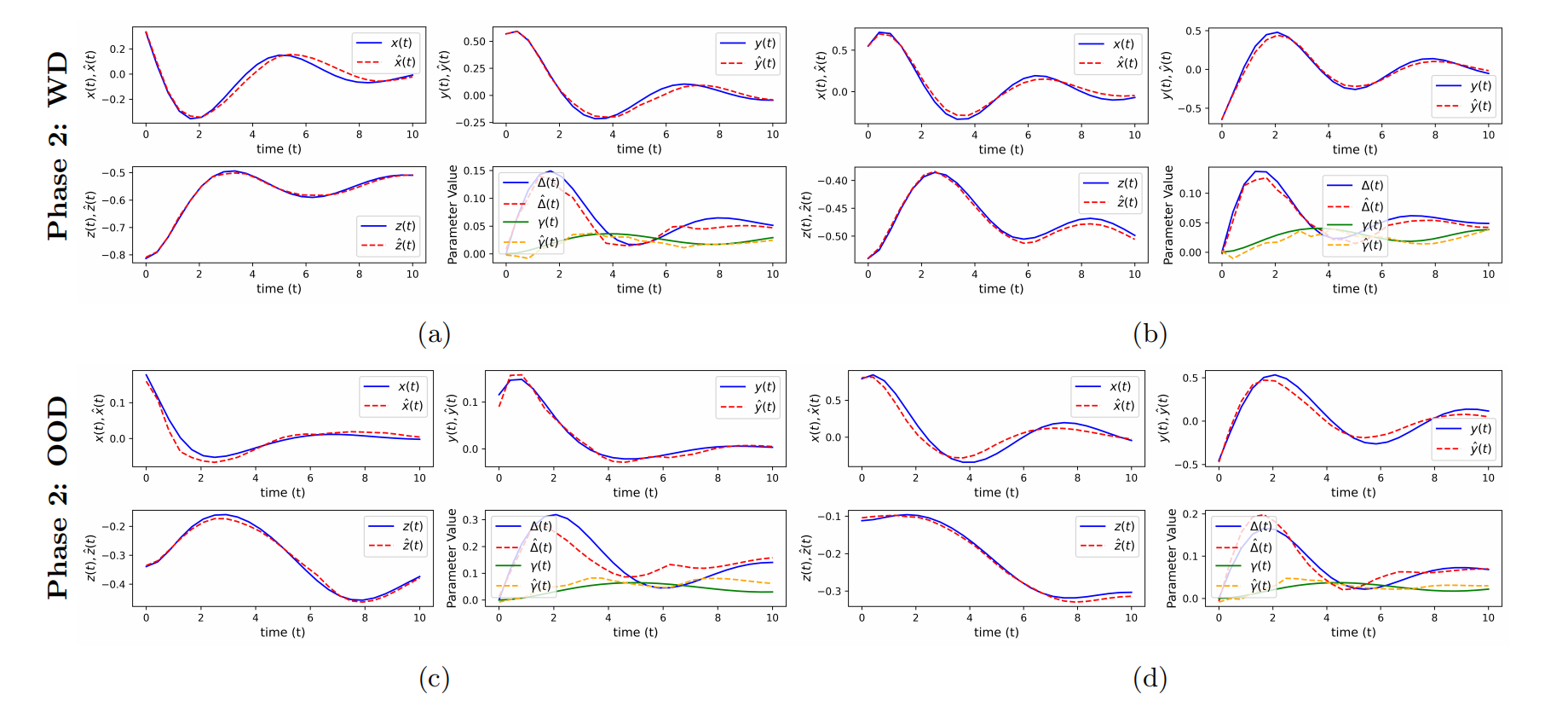}
    \caption{\justifying AQNODE prediction of qubit states and parameters in Phase 2 under WD (a–b) and OOD (c–d) conditions. The model is evaluated under variations in environmental parameters \( r \), \( \alpha \) (governing \( \Delta(t) \) and \( \gamma(t) \)), and system parameters \( \omega_0 \) and \( M \). In the WD regime (a–b), AQNODE shows accurate predictions of the qubit state \( x(t), y(t), z(t) \) and hidden parameters \( \Delta(t), \gamma(t) \). In the OOD regime (c–d), the model generalizes well, demonstrating its effectiveness in predicting the system’s evolution beyond the training distribution.}\label{Fig.7}
\end{figure*}

\begin{figure}[t]
    \centering
    \includegraphics[width=0.4\textwidth]{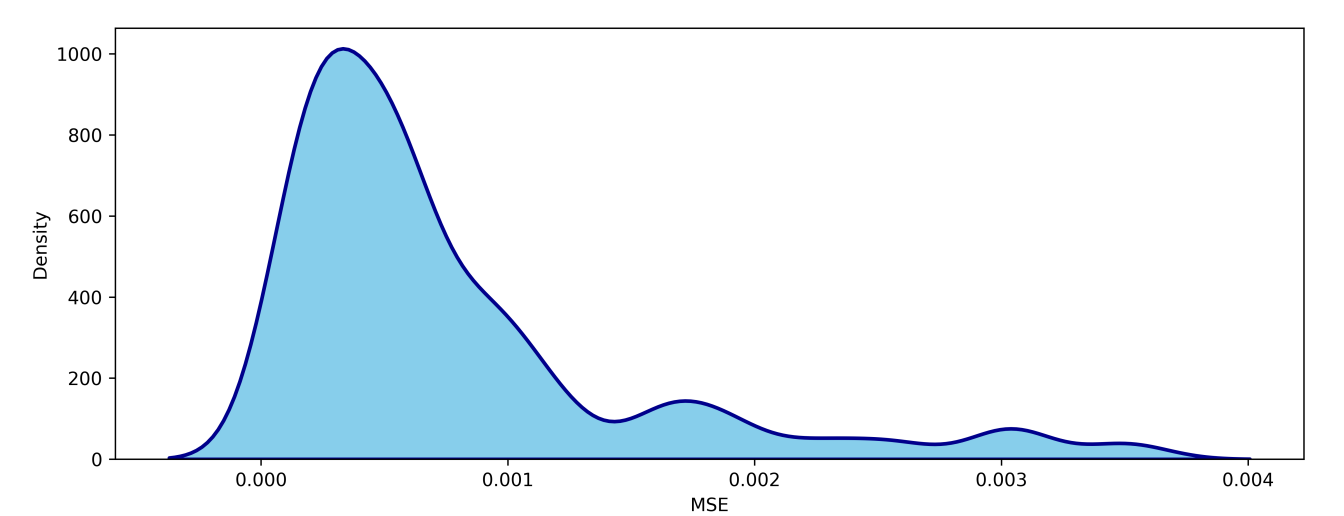}
    \caption{\justifying Phase 2: Average MSE over time.  The distribution exhibits a pronounced peak around MSE \(4\times10^{-3}\), the MSE distribution exhibits a right-skewed tail, indicating challenging outliers with higher error rates.\label{Fig.8}
}
\end{figure}

\begin{figure}[t]
    \centering
   \includegraphics[width=0.4\textwidth]{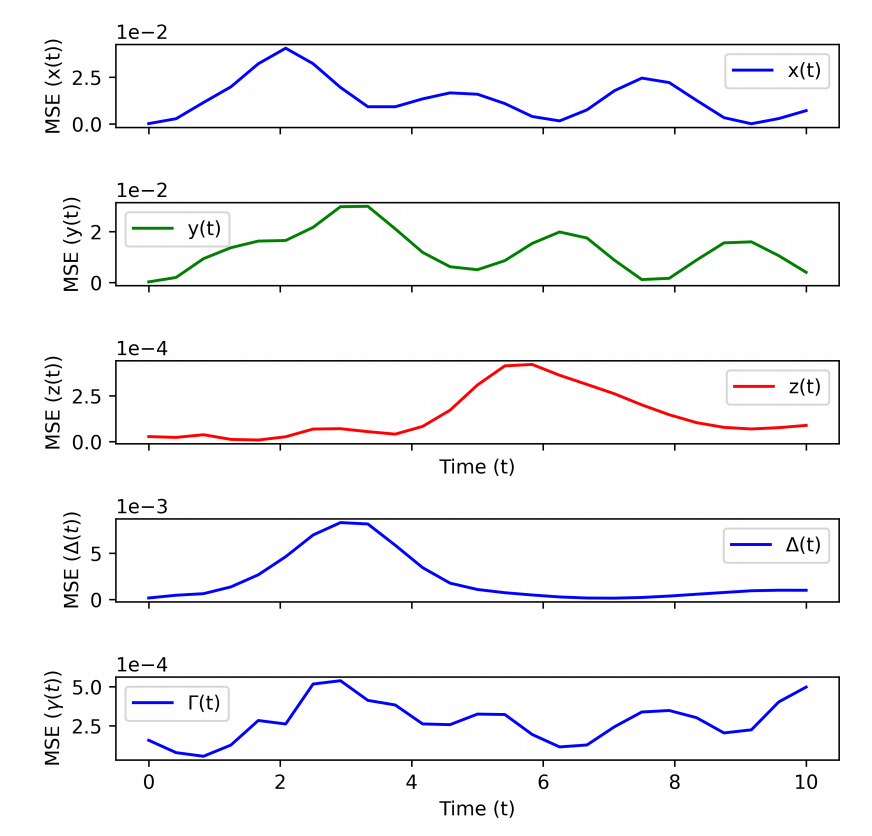}
    \caption{\justifying Average MSE curve over time during Phase 2, indicating consistent learning performance as AQNODE adapts to both system and environmental variability.}\label{Fig.9}
\end{figure}

\textit{Training Phase}: The model is trained under broader uncertainty, where both environmental and system parameters are randomized as follows $\alpha \in [0.4, 0.7],  \ r \in [0.2, 0.5],  \ M \in [0.3, 0.5], \ \text{and} \ \omega_0 \in [0.8, 1.2]$.

\textit{WD Testing}: 
The model is evaluated on new but slightly constrained distributions as $\alpha \in [0.45, 0.65], \ r \in [0.25, 0.45], \ M \in [0.32, 0.48] \ \text{and} \ \omega_0 \in [0.85, 1.15]$.
  
\textit{OOD Testing (Extreme Variations)}:  
  To check whether the model remains stable under completely unseen conditions, we allow extreme variability, i.e., $\alpha \in [0.2, 0.8], \ r \in [0.1, 0.6], \ M \in [0.2, 0.6] \ \text{and} \ \omega_0 \in [0.6, 1.5]$

This phase assesses how well AQNODE adapts when both the environment and system dynamics are unpredictable, testing its ability to function under realistic, dynamically changing quantum systems.

In Fig.~\ref{Fig.6}(a-e), shows the responses of Phase 2 during WD testing, the model is evaluated on parameter sets that fall slightly range as the training distribution, \(\alpha \in [0.4, 0.7], r \in [0.2, 0.5], M \in [0.3, 0.5]\), and \(\omega_0 \in [0.8, 1.2])\). The predicted trajectory overlaps with the real trajectory, confirming that the model has successfully learned the underlying quantum dynamics and can reconstruct unseen behaviors. In Fig.~\ref{Fig.6}(f-j) the OOD testing examines the model’s ability to generalize beyond its training range by evaluating trajectories with parameter values that extend beyond those seen during training. The key goal is to determine whether the learned AQNODE model can extrapolate correctly when exposed to new quantum dynamics. The model is tested on new quantum states that were not seen during training. This evaluates the extrapolation capability of the model. Again, the AQNODE model performed well on unseen cases.

The WD setting of phase 2 shows that the model can predict the time evolution of the Bloch vector components, as well as the dissipation and decoherence functions. The predicted values closely followed the true values, as evidenced in Fig.7 (a-d) shows the real and predicted results for \( x(t) \), \( y(t) \), \( z(t) \), \( \Delta(t) \), and \( \gamma(t) \)), both the predicted trajectory and the actual trajectory are close enough to predict the unknown state and parameters precisely. For the Bloch vector components and the parameters $\Delta(t)$ and $\gamma(t)$, the results offer a thorough comparison of the actual and anticipated trajectories for the OOD setting of phase 2, as depicted in Fig.~\ref{Fig.7}(a-d). Both transient and steady-state characteristics are accurately captured by the AQNODE, indicating its potential for quantum dynamics prediction and estimation. The MSE for the state and parameters and average training losses are shown in Figs.~\ref{Fig.8} and \ref{Fig.9}.

Moreover, the fact that AQNODE can estimate the state trajectory and governing parameters in the presence of uncertainty suggests that it functions not only as a predictive model but also as a quantum observer. That is, it extracts the most probable quantum state evolution from partial state observations. This model can be applied to real-time experiments to predict the state and parameters under noisy and incomplete measurements. 

\begin{figure}
\centering
\includegraphics[width=1\columnwidth]{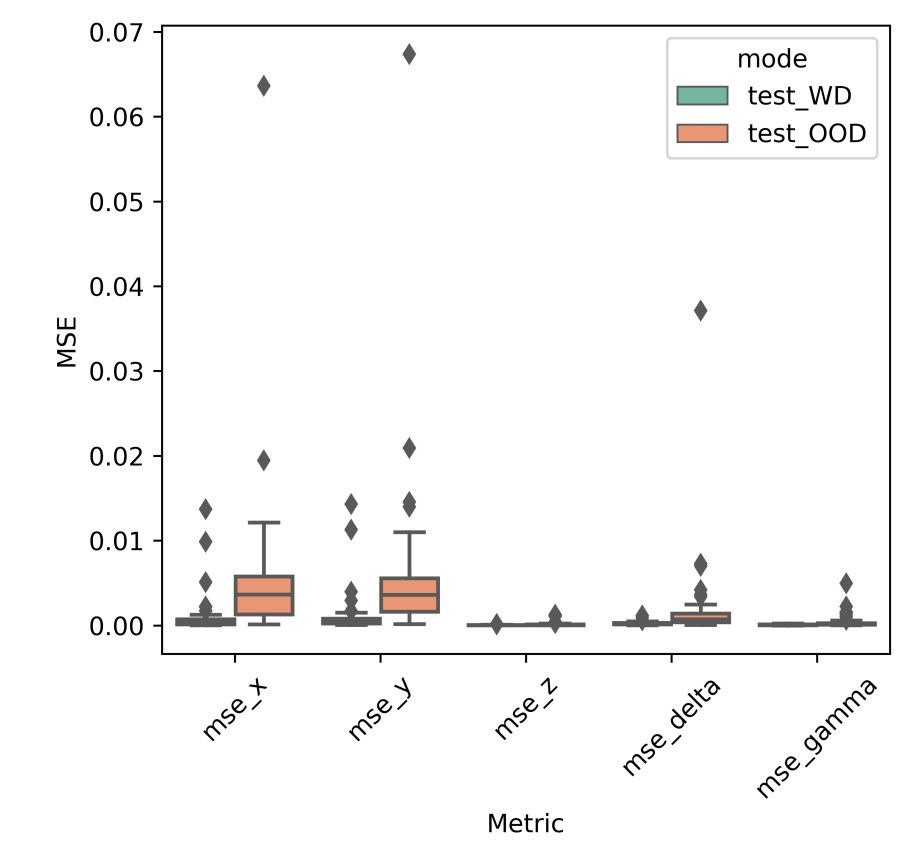}
\captionsetup{justification=justified}
\caption{\justifying Histogram of MSE distribution for Within-Distribution (WD) and Out-of-Distribution (OOD) test cases in Phase 2. The WD samples cluster tightly around lower error values, while the OOD distribution shows a broader tail, reflecting increased complexity and variability. The figure highlights AQNODE's generalization ability: while prediction accuracy slightly degrades under unseen conditions, the model remains stable and effective even with significant parameter shifts.}\label{Fig.10}
\end{figure}

The MSE of the state and the parameters are shown in Fig.~\ref{Fig.8}, and the average training loss and the MSE disturbation of WD and OOD is depicted in Fig.~\ref{Fig.9} and Fig.~\ref{Fig.10}.

\begin{table}[h]
\centering
\caption{\textbf{Phase 2:} Quantitative comparison of MSE for qubit state components and system parameters under WD and OOD test settings.}
\label{tab:mse_summary}
\begin{tabular}{lcc}
\textbf{Metric} & \quad \quad \textbf{WD} & \quad \quad \textbf{OOD} \\
\midrule
MSE~$x(t)$       & \quad \quad \( 1.08 \times 10^{-3} \) & \quad \quad \( 5.28  \times 10^{-3} \) \\
MSE~$y(t)$        & \quad \quad \( 1.16 \times 10^{-3} \) & \quad \quad \( 5.80 \times 10^{-2} \) \\
MSE~$z(t)$        & \quad \quad \( 5.60 \times 10^{-4} \) & \quad \quad \( 1.57 \times 10^{-3} \) \\
MSE~$\Delta(t)$   & \quad \quad \( 2.49 \times 10^{-4} \) & \quad \quad \( 1.99  \times 10^{-3} \) \\
MSE~$\gamma(t)$   & \quad \quad \( 1.02 \times 10^{-4}   \) & \quad \quad \( 4.34 \times 10^{-4} \) \\
\bottomrule
\end{tabular}
\label{tab:Phase-2-wd-ood}
\end{table}
In table \ref{tab:Phase-2-wd-ood}, the results show consistently low errors under WD, confirming accurate model predictions when tested on distributions similar to the training data. In contrast, higher MSEs under OOD, especially for $y(t)$ and $\Delta(t)$, reflect expected generalization limitations. However, the model retains reasonable predictive capacity, demonstrating robustness even under distributional change.

\textit{AQNODE prediction under initial perturbations}: 
To further evaluate the robustness and filtering capabilities of the AQNODE framework under non-Markovian quantum dynamics, we conduct an additional study in Phase 2 that investigates the model's response to perturbations in the initial quantum state to verify the AQNODE model's robustness against further uncertain changes.

We introduced perturbations to the 5-dimensional initial state vector $\vec{\mathcal{Y}}_0 = [x_0, y_0, z_0, \Delta_0, \gamma_0]$ as:
\[
\vec{\mathcal{Y}}_0^{( \epsilon)} = \vec{\mathcal{Y}}_0 + \epsilon \cdot {v} \ ,\quad \epsilon \in \{0.05, 0.1, 0.3\}
\]
where \(v\) is generated using a standard normal random vector \(n\), i.e., \({v} = \frac{{n}}{\|{n}\|},\quad {n}\sim \mathcal{N}(0, I) \). Each perturbed state is encoded by the AQNODE encoder, evolved via the trained neural ODE, and decoded into physical observables. We also compare AQNODE predictions under perturbed initial states with $\epsilon = 0.1$, $0.3$, and $0.5$, which simulate increasing uncertainty in the input. These perturbations are applied along a normalized direction in state space and represent AQNODE's response in perturbed filtering scenarios.

\begin{figure}[t]
    \centering
    \includegraphics[width=0.95\linewidth]{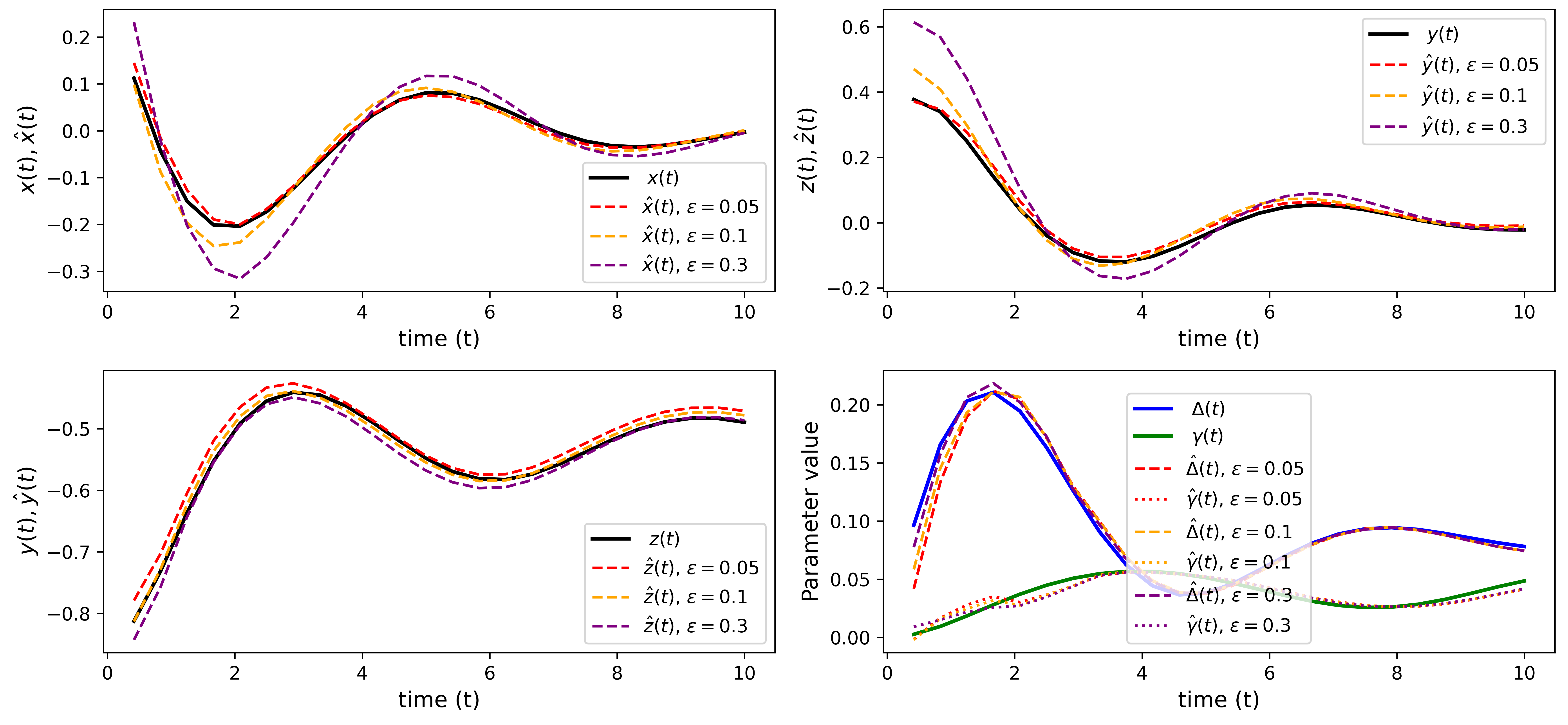}
    \caption{\justifying 
    AQNODE predictions under initial perturbations: $\hat{x}(t)$, $\hat{y}(t)$, $\hat{z}(t)$, $\hat{\Delta}(t)$, and $\hat{\gamma}(t)$, respectively, across perturbation magnitudes $\epsilon = 0.05$, $0.1$, and $0.3$. Solid lines denote ground truth and dashed lines indicate predictions under perturbed initial states. The AQNODE model exhibits smooth, bounded deviations and eventual convergence toward the true dynamics, consistent with quantum filtering behavior.}
    \label{Fig.11}
\end{figure}

Figure~\ref{Fig.11} presents four subplots that compare the ground truth $x(t)$, $y(t)$, $z(t)$ and the predicted trajectories of AQNODE $\hat{x}(t)$, $\hat{y}(t)$, $\hat{z}(t)$, respectively, along with the evolution of hidden dissipative parameters $\Delta(t)$, $\gamma(t)$ and their estimates $\hat{\Delta}(t)$, $\hat{\gamma}(t)$ under different initial perturbations. It is shown that the proposed AQNODE, under initial perturbations, deviates slightly in the early phase but ultimately follows the ground truth envelope. In particular, the hidden variables $\Delta(t)$ and $\gamma(t)$ are accurately inferred even under perturbed input states, although they are not explicitly measured, validating AQNODE's capacity to reconstruct latent quantum information.

These findings reflect the hallmark behavior of a quantum filter: Despite perturbations in the input state, the model leverages the measurement-informed latent representation to correct its prediction over time. The structure-aware dynamics learned by AQNODE enable this behavior and indicate that the model implicitly discovers a low-dimensional, physically consistent manifold on which the quantum trajectories evolve. This confirms that AQNODE acts as a data-driven quantum observer, exhibiting robust trajectory prediction under initial state uncertainty. The ability to recover both observable and latent dynamics from incomplete data strengthens AQNODE's candidacy as a real-time quantum state reconstruction framework, suitable for use in experimental feedback and estimation scenarios.


This behavior represents how the model is trained: at each time step, it must match its predictions to the actual system trajectory. The AQNODE learns to predict the dynamics even when the initial condition is perturbed.  The simulations show that AQNODE is capable of generalizing to unseen trajectories and maintaining stability despite initial shifts.

\subsection{Phase 3: Control of Dissipative Qubit Dynamics Using AQNODE Predictions}

The control of qubit dynamics is fundamental to quantum information processing, where precise manipulation of quantum states enables coherent computation, error correction, and quantum communication~\cite{edwards2005optimal,langen2024quantum,ryan2021realization}. We now turn to phase 3, in which the AQNODE-estimated state is used as the feedback variable for control. In this sense, the controller is not independent of the modeling stage; it is the downstream consumer of the latent-state inference produced by AQNODE. Our objective is to transfer the initial state to a desired final state \({\mathcal{Y}}_{\text{target}} = [0, 0, 1]^T\) while minimizing energy usage and preserving quantum state purity in the presence of environmental decoherence.  Crucially, the state \({\mathcal{Y}}(t) \in \mathbb{R}^3\) in  Eqn. (\ref{bloch_eqn}) used for feedback is not directly observed, but is inferred from a learned AQNODE model, trained to predict the qubit’s evolution based on indirect measurement signals and applied control fields. These control paradigms are used independently within the AQNODE framework to evaluate its flexibility and reliability in directing the system to the intended target state.  We illustrate AQNODE's ability to generalize over unknown configurations while preserving robust trajectory prediction and efficient closed-loop control by evaluating the prediction and control steps simultaneously.

The control-based neural ODE function (\(f_\theta(\mathcal{Y}_{\text{aug}}(t), dY(t), {u}(t))\)) has 34,309 parameters, while the decoder has 17,667 parameters. Together, these components make up the 51,976 trainable parameters of the neural ODE model. The parameter counts refer to trainable neural-network weights and biases denoted by $\psi, \theta,\phi$, not to physical parameters of the qubit. These two networks have different parameter sets because they perform different tasks: the Neural ODE vector field learns the continuous-time evolution of the internal state estimate, whereas the decoder maps this internal state to the physical variables. The number of trainable parameters is determined by the selected input dimension, output dimension, hidden width, number of layers, and bias terms. For a fully connected multilayer perceptron with layer widths $n_0,n_1,n_2,...,n_L$, the total number of trainable parameters is
$N_p=\sum_{l=1}^L(n_{l-1}+1)n_l$, where the additional 1 accounts for the bias in each layer.

The architecture was selected empirically to balance accuracy, generalization, computational cost, and numerical stability of the ODE solver. Reducing the number of parameters decreases model capacity  and increasing the number of parameters beyond the selected architecture produced only marginal improvement in prediction accuracy while increasing computational cost and, in some cases, reducing the smoothness and numerical stability of the learned ODE dynamics. Thus, the selected architecture should be understood as a numerical model-selection choice rather than a fundamental physical parameter. With these settings, ODE integration remains cheap and stable, and performance improves on both WD and OOD profiles. We chose them at the knee of the trade-off curve, where accuracy has plateaued while solver stability stays robust.

\subsubsection{LQR Control}
To achieve optimal control of the quantum state, LQR control dynamically adjusts feedback gains based on a cost function~\cite{doherty1999feedback,edwards2005optimal}. The evolution of the Bloch vector is governed by the \( \hat{\mathcal{Y}}(t)= [\hat{x}(t), \hat{y}(t), \hat{z}(t)]^T\), and the control fields \(u_x(t), u_y(t)\) are applied via the Hamiltonian operators \(A_x\) and \(A_y\). Thus, the control matrix \(B(t)\) is constructed as:

\begin{equation}
B = \begin{bmatrix} A_x {x}_{\text{target}} & A_y{x}_{\text{target}} \end{bmatrix}
\end{equation}

where \(\mathcal{Y}_{\text{target}} = [0, 0, 1]^T\) is the desired Bloch state.

The cost function has the form
\begin{eqnarray}
\begin{split}
&J({\hat{\mathcal{Y}}(t)}, \mathcal{Y}_{\text{target}}, {u(t)})=\\& \int_0^T \left[ (\hat{\mathcal{Y}}(t) - \mathcal{Y}_{\text{target}})^T Q (\hat{\mathcal{Y}}(t) - \mathcal{Y}_{\text{target}})\right]dt +\\& \int_0^T \left[{u}(t)^T R {u}(t) \right] dt
\end{split}
\end{eqnarray}
where \(Q \in \mathbb{R}^{3 \times 3}\) penalizes state deviations and \(R \in \mathbb{R}^{2 \times 2}\) penalizes control energy. In our system settings we define $Q = \text{diag}[1000, 1000, 1000]$ and $R = \text{diag}[0.1, 50]$. To obtain the optimal time-varying control law, we have

\begin{equation}
{u}^*(t) = -K(t) (\hat{\mathcal{Y}}(t) - {\mathcal{Y}}_{\text{target}})
\end{equation}
we compute the gain matrix \(K(t)\) by solving the differential Riccati equation

\begin{equation}
\frac{d}{dt} P(t) = -A(t)^T P(t) - P(t) A(t) + P(t) B R^{-1} B^T P(t) - Q
\end{equation}
where the optimal feedback gain $K(t)$ is then

\begin{equation}
K(t) = R^{-1} B^T P(t)
\end{equation}

This backward integration ensures that we accommodate the time-dependence of the quantum system due to decoherence. This structure allows flexible and optimal control over a quantum system under dissipation and ensures the qubit is steered close to the target quantum state efficiently.

\subsubsection{Phase 3: PD control}

\begin{figure*}[t]
    \centering
    \includegraphics[width=0.75\textwidth]{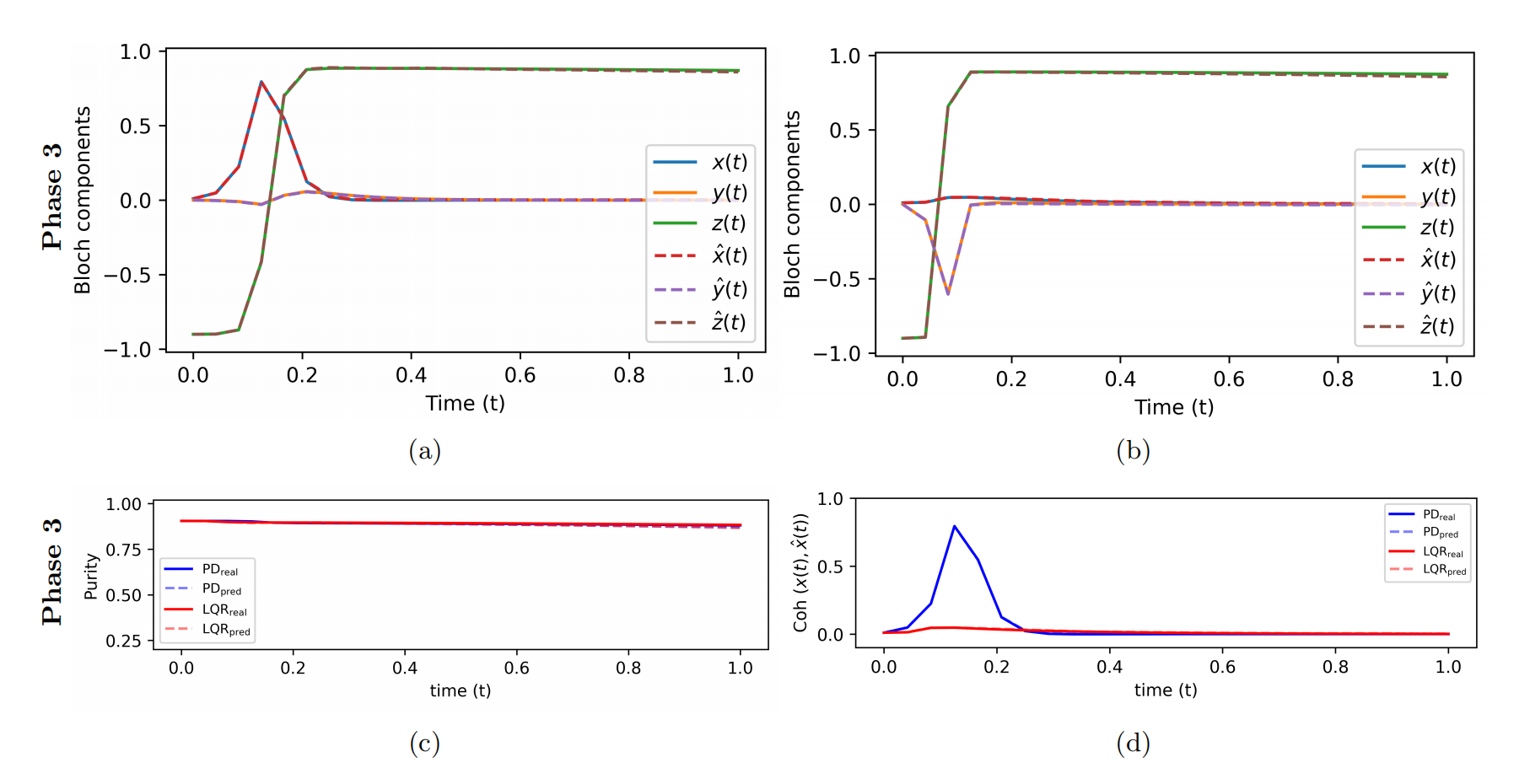}
    \caption{\justifying \textbf{WD regime:} Neural ODE-based prediction of qubit trajectories and purity under PD and LQR control in a WD setting. (a) and (b) show the Bloch vector components \((x(t), y(t), z(t))\) for PD and LQR controllers, respectively, to steer the system toward the target eigenstate \(|0\rangle\), with predictions closely matching the actual dynamics. (c) compares quantum purity over time for real and predicted trajectories. (d) shows coherence evolution along the \(x\)-axis, highlighting transient oscillations under PD control and rapid convergence under LQR control.}\label{Fig.12}
\end{figure*}

\begin{figure}[t]
    \centering
    \includegraphics[width=0.5\textwidth]{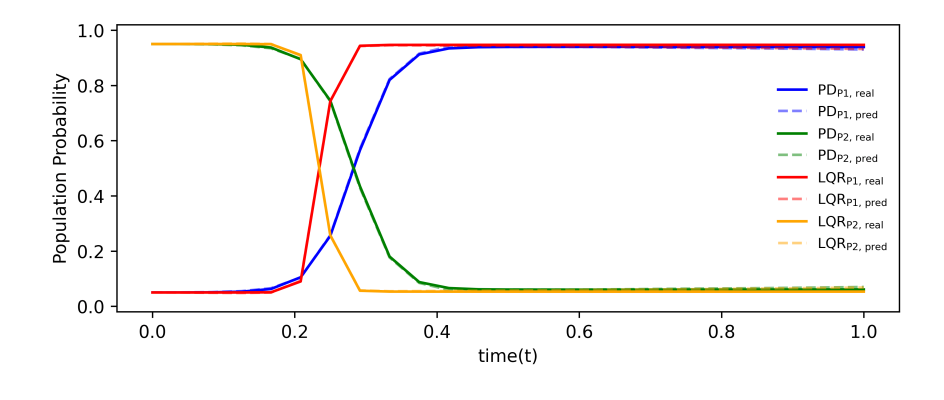}
    \caption{\justifying Time evolution of population probabilities $P_1(t)$ and $P_2(t)$ for real (solid) and predicted (dashed) dynamics under PD and LQR feedback control strategies in the WD setting. AQNODE accurately models the population transfer dynamics and timing under both control strategies.}\label{Fig.13}
\end{figure}

\begin{figure*}[t]
    \centering
    \includegraphics[width=1\textwidth]{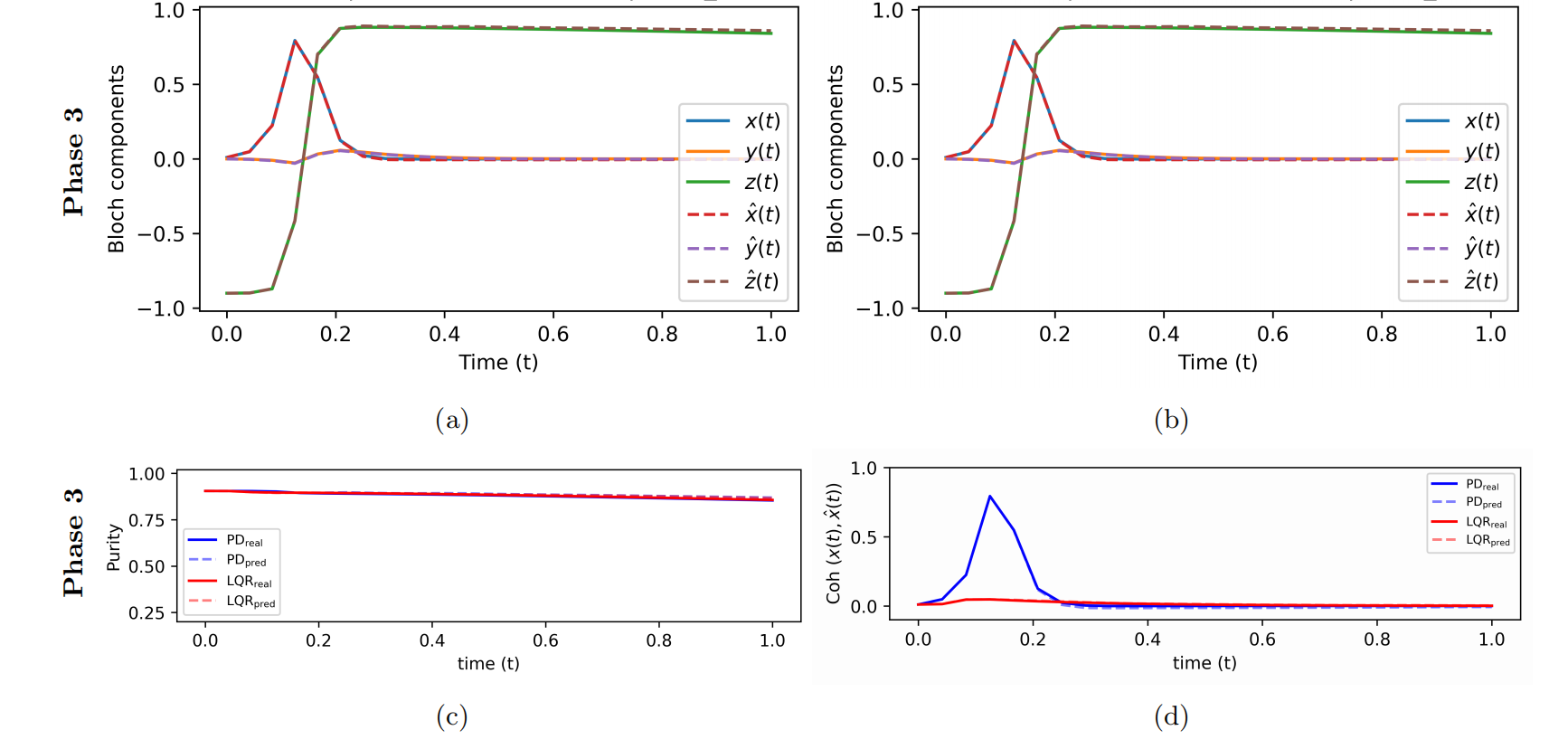}
    \caption{\justifying \textbf{OOD regime:} AQNODE-based prediction of qubit dynamics under PD and LQR control in the OOD setting. (a) and (b) show Bloch vector components \((x(t), y(t), z(t))\) for PD and LQR controllers. (c) compares purity, indicating precise prediction and sustained coherence, and (d) shows coherence.}\label{Fig.14}
\end{figure*}

\begin{figure}[t]
    \centering
    \includegraphics[width=0.5\textwidth]{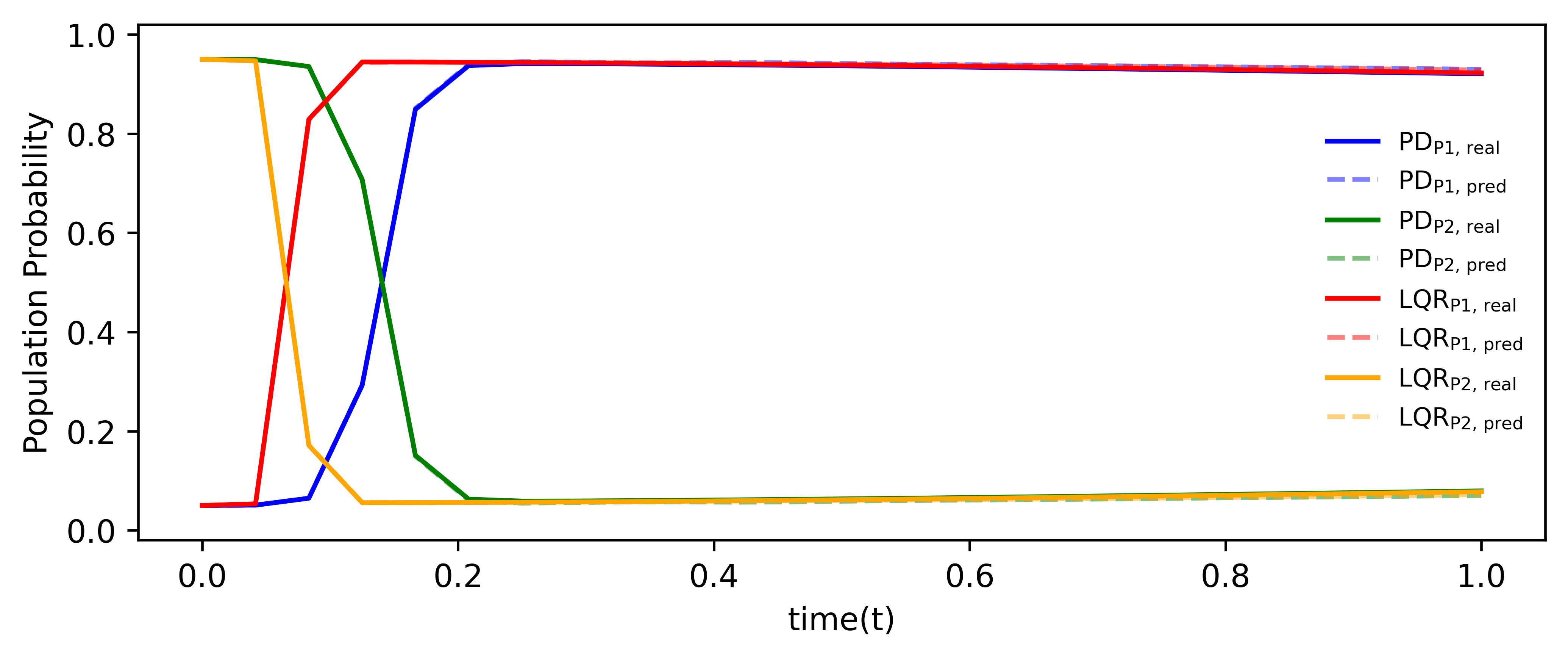}
    \caption{\justifying Time evolution of population probabilities $P_1(t)$ and $P_2(t)$ for real (solid) and predicted (dashed) dynamics under PD and LQR control paradigms in the OOD setting.}\label{Fig.15}
\end{figure}

\begin{figure}[t]
    \centering
    \includegraphics[width=0.5\textwidth]{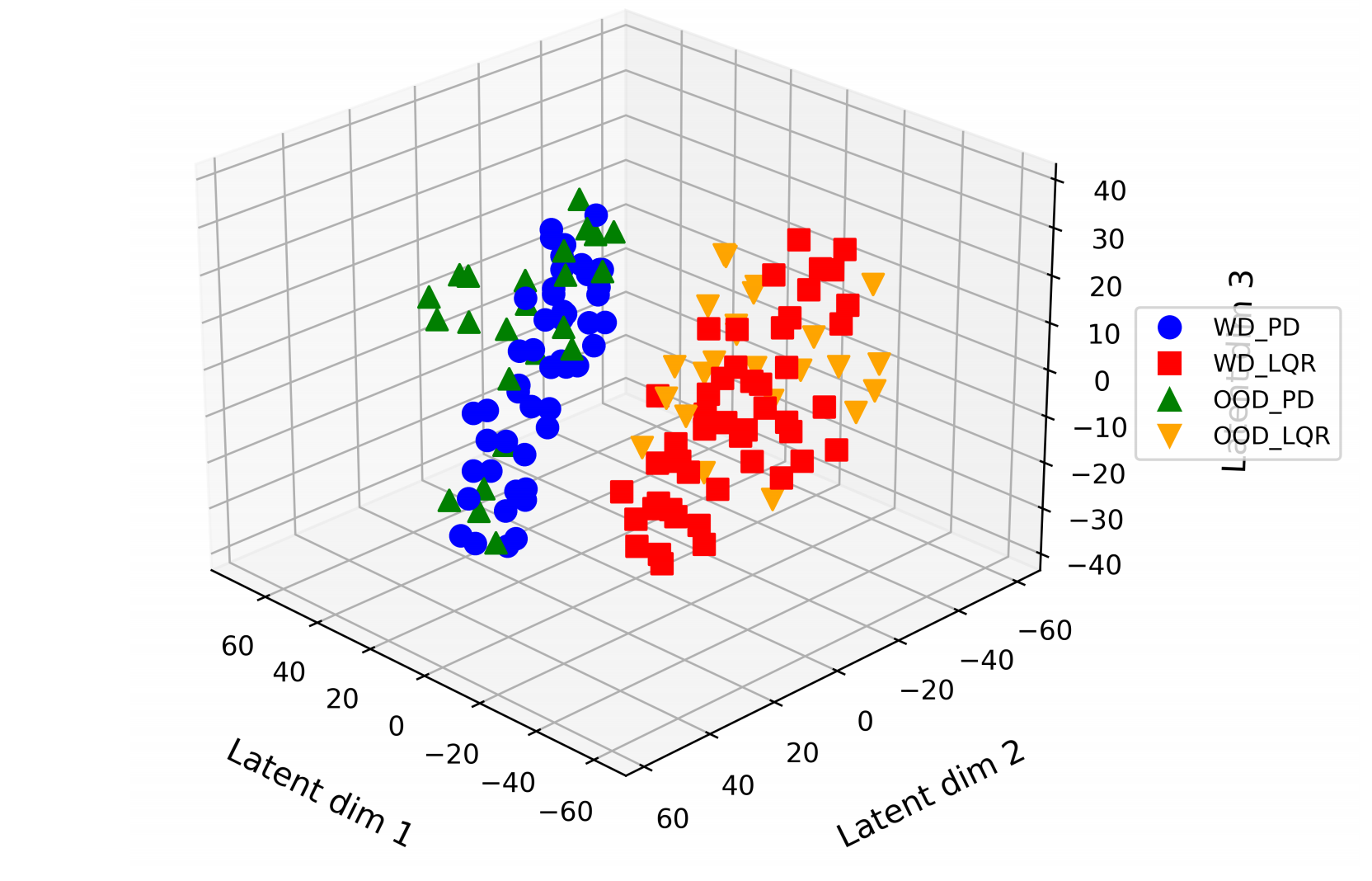}
    \caption{\justifying Visualization of the encoded latent space for test trajectories under different control strategies and distribution settings. Trajectories encoded by AQNODE are shown, with blue and red representing WD data using PD and LQR controls, and green and orange representing OOD samples. The clear clustering by control type and distribution demonstrates AQNODE’s ability to capture underlying dynamics, with OOD trajectories remaining well-separated and coherent, demonstrating its generalization capability.}\label{Fig.16}
\end{figure}

The PD control strategy computes the control fields based on the error \((\hat{\mathcal{Y}}(t) - \mathcal{Y}_{\text{target}})\) and its derivative \(\dot{\hat{\mathcal{Y}}}(t)\)~\cite{chen2020quantum, gough2020quantum}. Then the control fields are calculated as
\begin{eqnarray}
u_x = -k_p^x \cdot e_x (t) - k_d^x \cdot \dot{\hat{x}}(t)\\
u_y = -k_p^y \cdot e_y (t) - k_d^y \cdot \dot{\hat{y}}(t)
\end{eqnarray}
The errors \(e_x (t)=\hat{x}(t)-x_{target}(t)\) and \(e_y (t)=\hat{y}(t)-y_{target}(t)\) between the estimated current and desired states along \( \hat{x}(t) \) and \( \hat{y}(t) \) are represented by \( e_x(t) \) and \( e_y(t) \), respectively. The momentum of the system along these axes is represented by the time derivatives of the states \( \dot{x}(t) \) and \( \dot{y}(t) \). The derivative gains (\( k_d^x, k_d^y \)) take into account the system's velocity to avoid overshooting, while the proportional gains (\( k_p^x, k_p^y \)) guaranty the controller reacts to the amount of the error. In our system settings, we set $k_p^x; k_p^y = [5; 10]$ and $k_d^x; k_d^y = [8;10]$. Algorithm 2 summarizes the closed-loop control workflow used in Phase 3, where AQNODE-estimated states are used to compute PD and time-varying LQR feedback controls.
\begin{algorithm}[H]
\caption{State transfer of qubit based on AQNODE predictions}
\label{alg:control_combined}
\begin{algorithmic}[1]
\Require Estimated state $\hat{\mathcal{Y}}(t_k)$ from AQNODE; target state $\mathcal{Y}_{\mathrm{target}}$;
controller type $\in\{\mathrm{PD},\mathrm{LQR}\}$.
\Ensure Control fields $u_x(t_k),u_y(t_k)$.

\If{controller = PD}
  \For{$k=1,\dots,N$}
    \State Compute $e_x(t_k),e_y(t_k)$ and $\dot{\hat{x}}(t_k),\dot{\hat{y}}(t_k)$.
    \State Set
    \[
    u_x(t_k)\leftarrow -k_p^x e_x(t_k)-k_d^x \dot{\hat{x}}(t_k),\]
    \[u_y(t_k)\leftarrow -k_p^y e_y(t_k)-k_d^y \dot{\hat{y}}(t_k).
    \]
  \EndFor
\Else
  \State Solve the Riccati equation backward to obtain $P(t_k)$.
  \For{$k=0,\dots,N$}
    \State Compute $K(t_k)\leftarrow R^{-1}B^T(t_k)P(t_k)$.
    \State Set
    \[
    u(t_k)\leftarrow -K(t_k)\big(\hat{\mathcal{Y}}(t_k)-\mathcal{Y}_{\mathrm{target}}\big),
    \]
    \[
    u_x(t_k)\leftarrow [u(t_k)]_1,
    \qquad
    u_y(t_k)\leftarrow [u(t_k)]_2.
    \]
  \EndFor
\EndIf
\end{algorithmic}
\end{algorithm}
\subsubsection{Preliminaries of state transfer}
Coherence and purity are essential markers of the dynamics of quantum systems. Pure (\(p(t) = 1\)) and mixed (\(p(t) < 1\)) quantum states are distinguished by purity, i.e., \( p(t) = \frac{1 + x^2(t) + y^2(t) + z^2(t)}{2} \), which reflects the degree of decoherence or dissipation in the system. Its temporal evolution shows how energy splitting (\(\Delta(t)\)) and dissipation (\(\gamma(t)\)) either increase purity or push the system towards a mixed state (\(p(t) < 1\)). Coherence, denoted by \(x(t)\), reflects quantum superposition through the off-diagonal elements of the density matrix. It has a direct impact on purity because purity usually increases with coherence. The control fields adjust \(x(t)\) and \(z(t)\) as the system evolves, balancing energy splitting and dissipation to maintain purity close to 1 and coherence close to a desired value. Occupation probabilities represent the population of the system's eigenstates, \(P_1 = \frac{1 + x_3(t)}{2}\) and \(P_2 = \frac{1 - x_3(t)}{2}\). Control fields and non-Markovian effects drive the system's state transfer during evolution, which is reflected in these complementary probabilities (\(P_1 + P_2 = 1\)). The combined purity, coherence, and occupancy probabilities reveal the interplay between control, dissipation, and state stability, which provides a comprehensive picture of the trajectory of the quantum system.

The system initialize at \({X}_0 = [0. 0, -1.0]\) and the target is \({X}_{target} = [0, 0, 1]\) indicating maximum population in the $\ket{0}$ state.
The natural dynamics (drift) tend to reduce coherence and purity due to decay and detuning. The control fields counteract these effects to steer the system toward the target state. Figure~\ref{Fig.12} provides a comprehensive evaluation of the AQNODE-based model’s prediction and control fidelity for quantum state evolution within the WD. Figure~\ref{Fig.12}(a) illustrates the real and predicted Bloch vector components under PD control. Here, the PD controller induces a strong, transient oscillation, as evidenced by the sharp peak around \(t \approx 0.15\), and quickly stabilizes the state toward the target \([0, 0, 1]\). The AQNODE-based LQR control model successfully captures both the transient dynamics and the final stabilization, as seen in the close alignment of solid and dashed curves. Figure~\ref{Fig.12}(b) shows the corresponding trajectories under real and AQNODE-based LQR control. Unlike PD, LQR enforces smoother and monotonic convergence to the target state, and the predicted trajectories closely follow the real system with minimal overshoot or deviation. Figure~\ref{Fig.12}(c) depicts the purity, which quantifies the degree of mixedness of the quantum state. The LQR controller achieves and maintains near-unity purity throughout the evolution, indicating minimal decoherence, while PD control momentarily reduces purity during the transient phase. The predicted purity trends (dashed lines) closely track the real values (solid lines), highlighting the model’s ability to reconstruct fine-grained dissipative effects and predict decoherence dynamics. Finally, Fig.~\ref {Fig.12}(d) reports a key observable coherence. The pronounced peak in the PD curve reflects strong feedback-induced oscillations, whereas the coherence remains flatter and higher under LQR, affirming its stability. The coherence profile aligns with expectations and confirms that the AQNODE model can accurately forecast time-resolved deviations and matches the real-time controlled dynamics. This validates the AQNODE model’s ability to predict quantum system evolution under feedback control, including not only the state trajectories but also control-sensitive properties like purity and coherence. The fidelity of predicted dynamics, particularly under LQR, suggests strong generalization to WD tasks and robustness of the learned representation over PD control.

Figure~\ref{Fig.13} illustrates the time evolution of the population probabilities \( P_1(t) \) and \( P_2(t) \) for a PD and LQR in the WD scenario. Under PD (blue and green lines), the transition toward the $\ket{0}$ state \( P_1(t) \) begins gradually. Around \( t \approx 0.1 \), the populations begin to diverge, and by \( t \approx 0.41 \), the state occupation shifts toward \( P_1(t) \approx 1 \), with \( P_2(t) \) correspondingly approaching zero. The predicted trajectories (dashed blue and dashed green) closely follow the real PD-controlled dynamics, with only minor deviations at intermediate time steps. In contrast, LQR (red and orange lines) achieves a much faster population transfer. The sharpest transitions occur between \( t = 0.09 \) and \( t = 0.28 \), where \( P_1(t) \) swiftly rises to near-unity, while \( P_2(t) \) drops sharply toward zero. This rapid convergence highlights the ability of LQR to reduce final-state deviation and produce smoother stabilization under the chosen quadratic cost. However, it requires larger total control energy than PD for the weights used here. 

Similarly, Fig~\ref{Fig.14}(a-d) shows that the AQNODE-based PD and LQR outperform under the OOD regime. The AQNODE model's predictions (dashed red and orange) track the real dynamics remarkably well across the entire time window, capturing both the steep transition region and the final steady-state behavior. Overall, the time-resolved population dynamics confirm that AQNODE faithfully reproduces both slow PD-driven and fast LQR-driven state transfer mechanisms. This shows that the AQNODE-based control performance remains consistent across early, mid, and late time intervals, reinforcing its reliability for time-dependent quantum state control modeling. The statistical analyses comparing real and predicted PD and LQR control trajectories under both WD and OOD settings, along with supplementary results demonstrating the AQNODE-based control performance in OOD scenarios, are provided.

\renewcommand{\arraystretch}{1.1}
\setlength{\tabcolsep}{3pt}
\begin{table}[h]
\centering
\caption{Control evaluation metrics under WD and OOD test settings.}
\label{tab:control-metrics-wd-ood}
\begin{tabular}{llcccc}

\textbf{Mode} & \textbf{Control} & \textbf{MSE}  & \textbf{Energy}  & \textbf{Dev} & \textbf{Fidelity} \\
\midrule
\multirow{4}{*}{WD}
    & Real PD   & $3 \times 10^{-5}$ & 1698.6  & 0.415 & 0.939 \\
    & Pred PD   & $2 \times 10^{-4}$ & 2310.9  & 0.419 & 0.930 \\
    & Real LQR  & $2 \times 10^{-5}$ & 4315.3  & 0.281 & 0.941 \\
    & Pred LQR  & $1 \times 10^{-4}$ & 5624.1  & 0.291 & 0.932 \\
\midrule
\multirow{4}{*}{OOD}
    & Real PD   & $5 \times 10^{-4}$ & 1769.4  & 0.426 & 0.924 \\
    & Pred PD   & $3 \times 10^{-3}$ & 2173.1  & 0.429 & 0.931 \\
    & Real LQR  & $2 \times 10^{-4}$ & 4419.4  & 0.292 & 0.926 \\
    & Pred LQR  & $1 \times 10^{-3}$ & 5724.5  & 0.296 & 0.933 \\
\bottomrule
\end{tabular}
\end{table}

The table \ref{tab:control-metrics-wd-ood} provides a comprehensive comparison of PD and LQR control strategies applied to qubit state transfer under both WD and OOD conditions, using both real and AQNODE predicted dynamics. Key evaluation metrics include MSE between predicted and true trajectories, total control energy \(E = \int_0^T (u_x^2 + u_y^2) dt\), final state deviation \(\|\hat{\mathcal{Y}}(T) - \mathcal{Y}_{\text{target}}\|^2\), and fidelity \(F = \frac{1}{2}(1 + \hat{\mathcal{Y}}(T) \cdot \mathcal{Y}_{\text{target}})\) with respect to the target Bloch vector. Across both the WD and OOD regimes, overall LQR consistently demonstrates superior performance over PD by achieving lower energy expenditure, reduced final-state deviation, and higher fidelity, reflecting more efficient and precise state transfer. Furthermore, AQNODE’s predicted control responses align strongly with real dynamics, particularly in LQR-controlled cases, highlighting the model’s generalization capability and robustness to distributional shifts in dissipation parameters.  The predicted controls from the AQNODE model closely match the real dynamics, with the predicted LQR maintaining high fidelity $(\geq 0.94)$ even under OOD conditions. This validates the integration of model-based LQR feedback with learned quantum dynamics for reliable and optimal qubit control. Figure~\ref{Fig.15} depicts the time evolution of the population probabilities \(P_1(t)\) and \(P_2(t)\) for a PD and LQR in
OOD regime. As the control parameters are not globally optimized, this may give a small lag in the fidelity of the qubit state in each case due to the random parameteric values. Fig.~\ref{Fig.16} visualizes the encoded latent space for the test trajectories under control strategies and distribution settings.

\section{Conclusion and Prospects}

In this work, we propose an AQNODE  framework to model quantum system dynamics, capable of learning and generating quantum trajectories from partial measurement data without an explicit dynamical model during inference. Our approach integrates AQNODEs with quantum measurement time series, allowing the system to reconstruct, predict, and extrapolate quantum trajectories with high precision. By incorporating weak measurement outputs, we demonstrate that AQNODE captures essential quantum dynamical properties while maintaining physical consistency. Remarkably, using measurement-driven learning, our model recovers key quantum properties, such as the state and non-Markovian environment parameters.

Furthermore, we extend AQNODE beyond single-qubit dynamics to open quantum systems with parameterized control inputs and decoherence effects. The inclusion of measurement data, along with time-dependent parameters \(\Delta(t)\) and \(\gamma(t)\), enables our framework to generalize across different system configurations, making it robust for tasks such as quantum state estimation and quantum control. Our numerical experiments showcase AQNODE's ability to generate physically valid quantum trajectories, even when extrapolating beyond training conditions. 

A key observation is that AQNODE maintains high prediction performance even under OOD conditions, where both the environmental parameters (\(\alpha, r\)) and the system parameters (\(M, \omega_0\)) vary significantly from the training regime. The model effectively captures the dissipative nature of the quantum system by learning an implicit representation of the underlying differential equations that govern the evolution of the state. This ability to infer missing information from measurement data alone aligns AQNODE with quantum filtering techniques, where incomplete knowledge of the system is compensated by iterative state estimation. In this sense, AQNODE functions as a data-driven quantum observer, enabling real-time tracking of quantum states under varying experimental conditions, making it highly relevant for applications in quantum sensing, feedback control, and adaptive quantum protocols. Furthermore, our research presents a fresh and potent paradigm that unites conventional quantum control techniques with data-driven machine learning approaches. We offer a reliable, flexible, and interpretable tool for modeling, regulating, and reconstructing the dynamics of open quantum systems by combining the AQNODE with PD and LQR control. This contribution has the potential to significantly advance quantum technology and make it easier to apply quantum systems in real-world settings.

Beyond the two-level quantum system, AQNODE can be applied to multi-qubit systems, though the challenge lies in generating sufficiently high-dimensional training data. This suggests potential applications in quantum error correction, variational quantum algorithms, and NISQ-era quantum simulations. Future extensions may leverage experimental data from superconducting qubits, incorporating randomized quantum circuits to improve AQNODE’s ability to model many-body quantum dynamics. By bridging machine learning and quantum mechanics, our framework moves towards data-driven quantum modeling, potentially enabling real-time observer-based feedback control and quantum-enhanced adaptive learning protocols. 

\section*{Acknowledgment}
This work is supported by the Shenzhen Fundamental Research Program (No.~JCYJ20250604145655074), and Shenzhen Key Laboratory of Advanced Functional Carbon Materials Research and Comprehensive Application (Grant No. ZDSYS20220527171407017).

\appendix
\section{Neural ODEs for Quantum System Evolution}

The time evolution of a quantum state is governed by an unknown function \( f \) that captures system dynamics and parameter evolution. To train the neural ODE efficiently, we use the adjoint sensitivity method. Let

\begin{equation}
\varrho_{aug}(t)=\begin{pmatrix}
h_O(t)\\
\theta\\
t
\end{pmatrix},
\end{equation}
whose dynamics are
\begin{equation}
\begin{aligned}
\frac{d \varrho_{aug}(t)}{dt}&=f_{\mathrm{aug}}(\varrho_{aug}(t))\\
&=\begin{pmatrix}f_{\theta}\big(h_O(t),dY(t), u(t),t\big)\\
0\\
1
\end{pmatrix}
\end{aligned}
\end{equation}

The augmented adjoint state is defined as
\begin{equation}
a_{\mathrm{aug}}(t)=\frac{\partial \mathcal{L}}{\partial \varrho_{aug}(t)}=
\begin{pmatrix}
a(t)\\
a_{\theta}(t)\\a_t(t)
\end{pmatrix}=
\begin{pmatrix}
\dfrac{\partial \mathcal{L}}{\partial h_O(t)}\\
\dfrac{\partial \mathcal{L}}{\partial \theta}\\
\dfrac{\partial \mathcal{L}}{\partial t}
\end{pmatrix}
\end{equation}
It satisfies the backward adjoint equation
\begin{equation}
\frac{d a_{\mathrm{aug}}^{T}(t)}{dt}=-a_{\mathrm{aug}}^{T}(t)\frac{\partial f_{\mathrm{aug}}}{\partial \rho_{aug}(t)}
\end{equation}

The Jacobian is
\begin{equation}
\frac{\partial f_{\mathrm{aug}}}{\partial \rho_{aug}(t)}=
\begin{pmatrix}\dfrac{\partial f_{\theta}}{\partial h_O} & \dfrac{\partial f_{\theta}}{\partial \theta} & \dfrac{\partial f_{\theta}}{\partial t}\\
0 & 0 & 0\\
0 & 0 & 0
\end{pmatrix}
\end{equation}

Therefore, the adjoint dynamics become
\begin{equation}
\left[\dot{a}^{T},\dot{a}_{\theta}^{T},\dot{a}_{t}\right]=-\left[a^{T}\frac{\partial f_{\theta}}{\partial h_O},a^{T}\frac{\partial f_{\theta}}{\partial \theta},a^{T}\frac{\partial f_{\theta}}{\partial t}\right]
\end{equation}
The backward integration starts from $t_N$ with terminal conditions
\begin{equation}
a(t_N)=\frac{\partial \mathcal{L}}{\partial h_O(t_N)},\qquad a_{\theta}(t_N)=0,
\end{equation}
and
\begin{equation}
a_t(t_N)=a^{T}(t_N)f_{\theta}\big(h_O(t_N),dY(t_N),t_N\big)\end{equation}

By integrating backward from $t_N$ to $t_0$, the gradient of the loss with respect to the trainable parameters is obtained as
\begin{equation}
\frac{d\mathcal{L}}{d\theta}=a_{\theta}(t_0)
\end{equation}
For sequential measurement data evaluated at multiple time points $t_n$, the adjoint variable is updated at each observation time by adding the corresponding local loss gradient. This allows AQNODE to compute parameter gradients efficiently while avoiding the storage of all intermediate ODE solver states.

\section{Non-Markovian Open Quantum System}

Quantum systems interacting with their surrounding environment exhibit decoherence, which leads to the loss of quantum coherence due to entanglement with the environment. To model such dynamics, a quantum system \( S \) embedded in a dissipative bath \( B \) and interacting with a time-dependent external control field \( H_c(t) \). The total Hamiltonian governing the system is given by:
\begin{equation}
H_{\text{total}} = H_S + H_B + H_{\text{SB}} + H_c(t),
\end{equation}

where \( H_S \) represents the free Hamiltonian of the system, \( H_B \) describes the Hamiltonian of the bath, \( H_{\text{SB}} \) defines the system-bath interaction responsible for decoherence, \( H_c(t) \) is the Hamiltonian associated with external control, which is time-dependent and acts to mitigate decoherence.

The control is represented as a linear combination of predefined control generators \(\{H_i\}\).
\begin{equation}
H_c(t) = \sum_i u_i(t) H_i,
\end{equation}
where \(H_i\) are Hermitian operators on \(\mathcal{H}_S\) and \(u_i(t)\) are real-valued control amplitudes.

The environment \(H_B\) is assumed to be a collection of independent harmonic oscillators with conjugate momenta \(p_k\), masses \(m_k\), frequencies \(\omega_k\), and coordinates \(x_k\):
\begin{equation}
H_B = \sum_{k}\left(\frac{p_k^2}{2m_k} + \frac{m_k\omega_k^2 x_k^2}{2}\right),
\end{equation}
To characterize the system-bath coupling, we assume a bilinear interaction:
\begin{equation}
H_{\text{SB}} = \alpha\sum_{n} A_{n} \otimes B_{n},
\end{equation}
where \( A_{n} \) and \( B_{n} \) are operators acting on the system and bath, respectively. In our two-level system, the assumed bilinear interaction
between the system $S$ and the environment $B$ can be written as
\begin{equation}
    H_{SB}=\alpha(\sigma^+ \otimes B+\sigma^- \otimes B^\dagger),\nonumber
\end{equation}
where $B=\sum_k g_k a_k$ and $B^\dagger=\sum_k g_k^*a_k^\dagger$, \(a_k\) and \(a_k^\dagger\) are the annihilation and creation operators of the \(k\)-th reservoir mode, and \(g_k\) is the coupling strength. Therefore, the system-side operators are \(A_1=\sigma^+\) and \(A_2=\sigma^-\), while the bath-side operators are \(B_1=B\) and \(B_2=B^\dagger\).
Transforming into the interaction picture, the Hamiltonian takes the form
\begin{equation}
\begin{aligned}
H_{\mathrm{SB}}(t)
&= e^{i(H_S+H_B)t}\,H_{\mathrm{SB}}\,
   e^{-i(H_S+H_B)t} \\
&= \alpha \sum_n A_n(t)\otimes B_n(t).
\end{aligned}
\end{equation}
where
\[
A_{n}(t) = e^{i H_S t} A_{n} e^{-i H_S t}, \quad B_{n}(t) = e^{i H_B t} B_{n} e^{-i H_B t}.
\]
Generally speaking, the interaction between the system and environment can be used to illustrate decoherence.
To describe the evolution of the system’s reduced density matrix \( \rho(t) \) is given by:
\[
\rho_S(t) = \text{Tr}_B [\rho_{\text{tot}}(t)],
\]
where \( \rho_{\text{tot}}(t) \) is the complete density matrix of the system and \(\text{Tr}_B\) denotes the partial trace over the bath \(\mathcal H_B\). In general, the state moves toward a statistical mixture and the off-diagonal parts of \(\rho_S(t)\) decay.  An initially pure state may change into a statistical mixture under decoherence.
\begin{equation}
\rho_S(t)\to \sum_n |c_n|^2 |a_n\rangle\langle a_n|
\end{equation}
in some basis \(\{|a_n\rangle\}\), which describes a statistical mixture of noninterfering states.  Therefore, analyzing the evolution of the nondiagonal elements of the reduced density matrix under the master equation is an often suggested method of decoherence analysis.

Considering the control in the presence of non-Markovian effects. The formal TCL equation for the relevant part \(\wp\rho(t)\) can be expressed as
\begin{equation}
\frac{d}{dt}\wp\rho(t)
=\sum_i u_i(t)\widetilde {\mathcal{T}}_i(t)\wp\rho(t)
+\mathcal{T}(t)\wp\rho(t)+\mathcal{D}(t)\mathcal {Q}\rho(t_0),
\end{equation}
with the time-local generators
\begin{equation}
\widetilde {\mathcal{T}}_i(t)=e^{iH_S t}H_i e^{-iH_S t},\qquad
\mathcal{T}(t)=\alpha\wp\mathcal L(t)[\mathbb 1-S(t)]^{-1}\wp
\end{equation}
and the inhomogeneity
\begin{equation}
\mathcal D(t)=\alpha\wp\mathcal L(t)[\mathbb{1}-S(t)]^{-1}g(t,t_0)\mathcal {Q}.
\end{equation}
where \(\wp\) is the projection superoperator on the relevant subspace and \(\wp\rho=\text{Tr}_B\{\rho\}\otimes\rho_B\). \(Q=\mathbb 1-\wp\) and \(\mathcal L(t)\) is the interaction-picture Liouvillian generated by system-bath interaction, the memory superoperator \(\mathcal S(t)\) has the form
\begin{equation}
 \mathcal S(t)=\alpha\int_{t_0}^{t}ds \mathcal G(t,s)\mathcal{Q}\mathcal L(s)\wp \mathcal G(t,s),
 \end{equation}
with \(\mathcal G\) the propagator in the interaction picture and \(g(t,t_0)\) encoding initial correlations. Eqns. (B7)–(B10) capture non-Markovian effects through the explicit time dependence of \(\mathcal T(t)\), the presence of \(\mathcal S(t)\), and the inhomogeneous term \(\mathcal D(t)\) (which vanishes for factorized initial conditions with \(\mathcal Q\rho(t_0)=0)\). To second order in the coupling constant, the general form of the master equation can be approximated by
\begin{equation}
\frac{d}{dt}\wp\rho(t)=\mathcal K(t)\wp \rho(t) + \mathcal D(t)\mathcal Q \rho(0)
\end{equation}
Equation (B7) makes explicit how control, dissipation, and initial system–bath correlations \(\mathcal{Q}\rho(t_0)\) jointly shape the evolution. One obtains a time-local generator \(\mathcal T(t)\) as in Eqns. (B8)–(B10). In general,
\begin{equation}
\begin{aligned}
\mathcal{T}(t)\rho_S(t)
=&-\frac{i}{\hbar}[H_S(t),\rho_S(t)]  \\
&+\sum_i \Big[
\mathcal{C}_i(t)\rho_S(t)\mathcal{E}_i^\dagger(t)
+\mathcal{E}_i(t)\rho_S(t)\mathcal{C}_i^\dagger(t)
\Big] \\
&-\frac{1}{2}\sum_i
\Big\{
\mathcal{E}_i^\dagger(t)\mathcal{C}_i(t)
+\mathcal{C}_i^\dagger(t)\mathcal{E}_i(t),
\rho_S(t)
\Big\}.
\end{aligned}
\end{equation}
where typically \(\mathcal C_i(t)\neq \mathcal E_i(t)\), which means that it is not in the Lindblad form.

According to the theory of open quantum systems, the dynamics in Eq.~(\ref{eqn:1}) describe a system coupled to an environment with finite memory. In the TCL formulation, these memory effects are represented through the time-dependent coefficients \(\Delta(t)\) and \(\gamma(t)\), rather than through an explicit memory-kernel integral. As a result, the equation remains time-local, while the coefficients carry information about the past system--environment interaction.

When the reservoir has a nontrivial spectral structure, information that has flowed from the system into the environment can partially return to the system at later times. This information backflow is a characteristic feature of non-Markovian dynamics. The parameters $\Delta(t)$ and $\gamma(t)$ up to second order in the system-reservoir coupling constant, are given by~\cite{cui2008optimal,PhysRevA.70.032113}
\begin{equation}\label{eqn:9}
\Delta \left( t \right)=\underset{0}{\overset{t}{\mathop \int }}\,d\tau k\left( \tau  \right)\text{cos}\left( {{\omega }_{0}}\tau  \right)
\end{equation}
\begin{equation}\label{eqn:10}
\gamma \left( t \right)=\underset{0}{\overset{t}{\mathop \int }}\,d\tau \mu \left( \tau  \right)\text{sin}\left( {{\omega }_{0}}\tau  \right)
\end{equation}
with
\begin{eqnarray}\label{eqn:11}\nonumber
k\left( \tau  \right)=2\underset{0}{\overset{\infty }{\mathop \int }}\,d\omega J\left( \omega  \right)\text{coth}\left[ \hbar \omega /2{{k}_{B}}T \right]\text{cos}\left( \omega \tau  \right)
\end{eqnarray}
\begin{eqnarray}\label{eqn:12}\nonumber
\mu \left( \tau  \right)=2\underset{0}{\overset{\infty }{\mathop \int }}\,d\omega J\left( \omega  \right)\text{sin}\left( \omega \tau  \right)
\end{eqnarray}
where $k_B T$ is the environmental temperature. $J(\omega)$ is the spectral density of the environment with Lorentz-Drude cutoff function
\begin{eqnarray}\label{eqn:13}
J\left( \omega  \right)=\frac{2}{\pi }\omega \frac{\omega _{c}^{2}}{\omega _{c}^{2}+{{\omega }^{2}}}
\end{eqnarray}
where $\omega$ is the bath frequency, and $\omega_c$ is the high-frequency cutoff. With an Ohmic Lorentz-Drude
cutoff spectral density $J(\omega)$ and finite cutoff $\omega_c$, the noise and dissipation kernels $k(\tau)$ and $\mu(\tau)$ determine the time-dependent coefficients using Eqns.~(B13)-(B14).
Finite $\omega_c$ induces non-exponential transients that encode bath memory; in the high-temperature limit these integrals reduce to the analytic expressions reported in Eqns.~(B16)-(B18), which we use in simulations.

Then the analytical expression in
Eq. (\ref{eqn:10}) can be expressed as
\begin{equation}
\gamma(t)
=
\frac{\alpha^2\omega_0 r^2}{1+r^2}
\bigg[
1-e^{-r\omega_0 t}
\bigg(
\cos(\omega_0 t)+r\sin(\omega_0 t)
\bigg)
\bigg],
\end{equation}
where $r$ is the ratio of $\omega_c$ and $\omega_0$.
\begin{figure}
    \centering
    \includegraphics[width=0.95\linewidth]{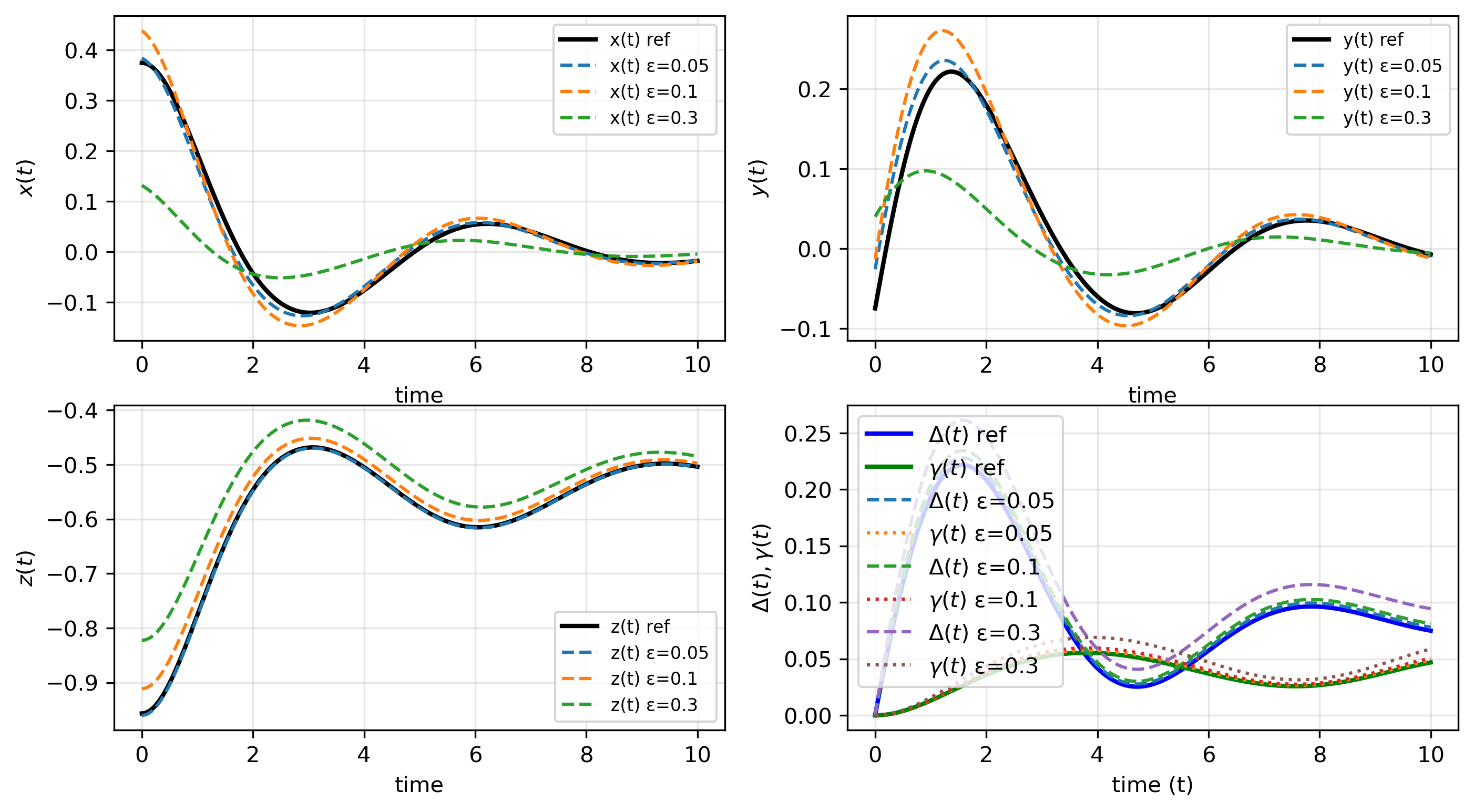}
    \caption{\justifying Comparison of perturbed and unperturbed trajectories of the qubit state and environmental parameters. The black solid curves indicate unperturbed initial conditions, while the dashed curves represent perturbed trajectories with different perturbations. It can be observed that the perturbations result in small deviations in the Bloch vector components.}
    \label{Fig.21}
\end{figure}
 The analytical expression of $\Delta \left( t \right)$ can be expressed as
\begin{multline}
\Delta(t) = \alpha^2 \omega_0 \frac{r^2}{1 + r^2} \bigg(
 \coth(\pi r_0) - \cot(\pi r_c) e^{-\omega_c t}\, [r \cos(\omega_0 t) \\
 - \sin(\omega_0 t)]
 + \frac{1}{\pi r_0} \cos(\omega_0 t)
   \left[ \bar{F}(-r_c,t) + \bar{F}(r_c,t) - \bar{F}(i r_0,t) \right. \\
   \left. - \bar{F}(-i r_0,t) \right]
 - \frac{1}{\pi r_0} \sin(\omega_0 t) \bigg\{
   \frac{e^{-v_1 t}}{2 r_0 (1 + r_0^2)} \left[ (r_0 - i)\,\bar{G}(-r_0,t) \right. \\
   \left. + (r_0 + i)\,\bar{G}(r_0,t) \right]
   + \frac{1}{2r_c} \left[ \bar{F}(-r_c,t) - \bar{F}(r_c,t) \right]
 \bigg\}
\bigg)
\end{multline}
where $r=\sfrac{\omega_c}{\omega_0}$, $r_0=\sfrac{\omega_0}{2\pi k_B T}$ and $r_c=\sfrac{\omega_c}{2\pi k_B T}$, in which $k_B T$ is the environmental temperature, and
\begin{eqnarray}\nonumber
\bar{F}\left( x,t \right)\equiv {}_{2}{{F}_{1}}\left( x,~1,~1,~1+x,{{e}^{-{{\nu}_{1}}t}} \right), \\
\bar{G}\left( x,t \right)\equiv {}_{2}{{F}_{1}}\left( 2,~1+x,2+x,~{{e}^{-{{\nu}_{1}}t}} \right)\nonumber
\end{eqnarray}
where \(\nu_1=2\pi k_B T\) and ${}_{2}{{F}_{1}}(a,b,c,z)$ is the Gauss hypergeometric function and is defined as
\begin{eqnarray}\nonumber
\begin{split}
{}_{2}{{F}_{1}}\left( a,~b,~c,~z \right) & =1+\frac{ab}{1!c}z+\frac{a\left( a+1 \right)b\left( b+1 \right)}{2!c\left( c+1 \right){{z}^{2}}}{{z}^{2}}+\ldots \\ \nonumber
& =\underset{n=0}{\overset{\infty }{\mathop \sum }}\,\frac{{{\left( a \right)}_{n}}{{\left( b \right)}_{n}}{{z}^{n}}}{{{\left( c \right)}_{n}}}\frac{{{z}^{n}}}{n!} \nonumber
\end{split}
\end{eqnarray}
where$ (a)_n$ is a Pochhammer symbol.
Under the high temperature limit, the diffusion coefficient $\Delta(t)$ now has the form
\begin{align} 
\Delta(t) &= 2\alpha^2 k_B T \frac{r^2}{1 + r^2} \left\{ 1 - e^{-r \omega_0 t} \left[ \cos(\omega_0 t) - \frac{1}{r} \sin(\omega_0 t) \right] \right\} \label{eqn:16}
\end{align}
The physical system itself demonstrates intrinsic stability, as seen in Fig.\ref{Fig.21}, where perturbations in the initial state diminish with time and the disturbed trajectories tend to converge toward the reference trajectory.  This demonstrates that a slight shift in the initial conditions do not significantly affect the system dynamics.
\bibliography{References}
\end{document}